\documentclass[prb,twocolumn,amsmath,amssymb,superscriptaddress,longbibliography]{revtex4-2}
\usepackage[colorlinks=true,citecolor=blue,linkcolor=blue,breaklinks=true]{hyperref}
\usepackage{amssymb}
\usepackage{graphicx}
\usepackage{amsmath}
\usepackage{comment}
\usepackage{mathtools}
\usepackage[export]{adjustbox}
\usepackage{epsfig}
\usepackage{times}
\usepackage{xcolor}
\usepackage{subfigure}
\usepackage{setspace}
\usepackage{bm}
\usepackage{calc}
\usepackage{natbib}
\usepackage[normalem]{ulem}
\usepackage{cancel}
\begin{document}

\definecolor{ao}{rgb}{0.0, 0.5, 0.0}

\newcommand{\fm}[1]{\textcolor{black}{#1}}
\newcommand {\ba} {\ensuremath{b^\dagger}}
\newcommand {\Ma} {\ensuremath{M^\dagger}}
\newcommand {\psia} {\ensuremath{\psi^\dagger}}
\newcommand {\psita} {\ensuremath{\tilde{\psi}^\dagger}}
\newcommand{\lp} {\ensuremath{{\lambda '}}}
\newcommand{\A} {\ensuremath{{\bf A}}}
\newcommand{\Q} {\ensuremath{{\bf Q}}}
\newcommand{\kk} {\ensuremath{{\bf k}}}
\newcommand{\qq} {\ensuremath{{\bf q}}}
\newcommand{\kp} {\ensuremath{{\bf k'}}}
\newcommand{\rr} {\ensuremath{{\bf r}}}
\newcommand{\rp} {\ensuremath{{\bf r'}}}
\newcommand {\ep} {\ensuremath{\epsilon}}
\newcommand{\nbr} {\ensuremath{\langle ij \rangle}}
\newcommand {\no} {\nonumber}
\newcommand{\up} {\ensuremath{\uparrow}}
\newcommand{\dn} {\ensuremath{\downarrow}}
\newcommand{\rcol} {\textcolor{red}}

\newcommand{\tblk}{\makebox[0pt] \boxed{~T~}}
\newcommand{\ac}[1]{\textcolor{magenta}{#1}}
\newcommand{\toremove}[1]{\textcolor{gray}{#1}}
\newcommand{\jhw}[1]{\textcolor{red}{#1}}
\newcommand{\jp}[1]{\textcolor{black}{#1}}
\newcommand{\JP}[1]{\textcolor{blue}{#1}}
\newcommand{\KC}[1]{\textcolor{green}{#1}}
\newcommand{\JW}[1]{\textcolor{purple}{#1}}

\begin{abstract}
We study critical properties of the entanglement and charge-sharpening measurement-induced phase transitions in a non-unitary quantum circuit evolving with a U(1) conserved charge. Our numerical estimation of the critical properties of the entanglement transition at finite system sizes appears distinct from the generic non-conserving case and percolation. We provide two possible interpretations of this observation: (a) these two transitions occur at different measurement rates in the thermodynamic limit, but at finite system sizes their critical fans overlap and the critical exponents we probed here show a combination of both the criticality.
Nonetheless, the multifractal properties 
of the entanglement transition remain distinct from the generic case without any symmetry, indicating a unique universality class due to the U(1) symmetry. (b) these two transitions occur at the same measurement rate at any length scale. Within this interpretation, our estimation of all the critical exponents are sharply different than the non-conserving case, again confirming the presence of a new universality class due U(1) symmetry.
We compute entanglement critical exponents and correlation functions via various ancilla measures, use a transfer matrix for multifractality, and compute correlators associated with charge sharpening to explain these findings.
Through these correlators, we also find evidence consistent with the charge-sharpening transition being of the Berezinskii-Kosterlitz-Thouless type (including the predicted ``jump'' in stiffness), 
which simultaneously argues for a broad critical fan for this transition.
As a result, attempts to measure critical properties in this finite-size system will see anomalously large exponents predicted by our numerical analysis.
 \end{abstract}

\title{Charge and Entanglement Criticality in a U(1)-Symmetric Hybrid Circuit of Qubits}
 \author{Ahana Chakraborty}\email{ahana@physics.rutgers.edu}
  \affiliation{Department of Physics and Astronomy, Center for Materials Theory,  
Rutgers University, Piscataway, NJ 08854, USA}
\affiliation{Center for Computational Quantum Physics, Flatiron Institute, 162 5th Avenue, New York, NY 10010, USA}
\author{Kun Chen}\email{kunchen@flatironinstitute.org}
\affiliation{Center for Computational Quantum Physics, Flatiron Institute, 162 5th Avenue, New York, NY 10010, USA}
\author{Aidan Zabalo}\email{zabalo@physics.rutgers.edu}
 \affiliation{Department of Physics and Astronomy, Center for Materials Theory,  
Rutgers University, Piscataway, NJ 08854, USA}
\author{Justin H. Wilson}\email{justin@jhwilson.com}
\affiliation{Department of Physics and Astronomy, Louisiana State University, Baton Rouge, Louisiana 70803, USA}
\affiliation{Center for Computation and Technology, Louisiana State University, Baton Rouge, Louisiana 70803, USA}
\author{J. H. Pixley}\email{jed.pixley@physics.rutgers.edu}
  \affiliation{Department of Physics and Astronomy, Center for Materials Theory,  
Rutgers University, Piscataway, NJ 08854, USA}
\affiliation{Center for Computational Quantum Physics, Flatiron Institute, 162 5th Avenue, New York, NY 10010, USA}

\date{\today}

\maketitle

\section{Introduction:}

\begin{figure*}[t!]
 \includegraphics[width=0.99\textwidth]{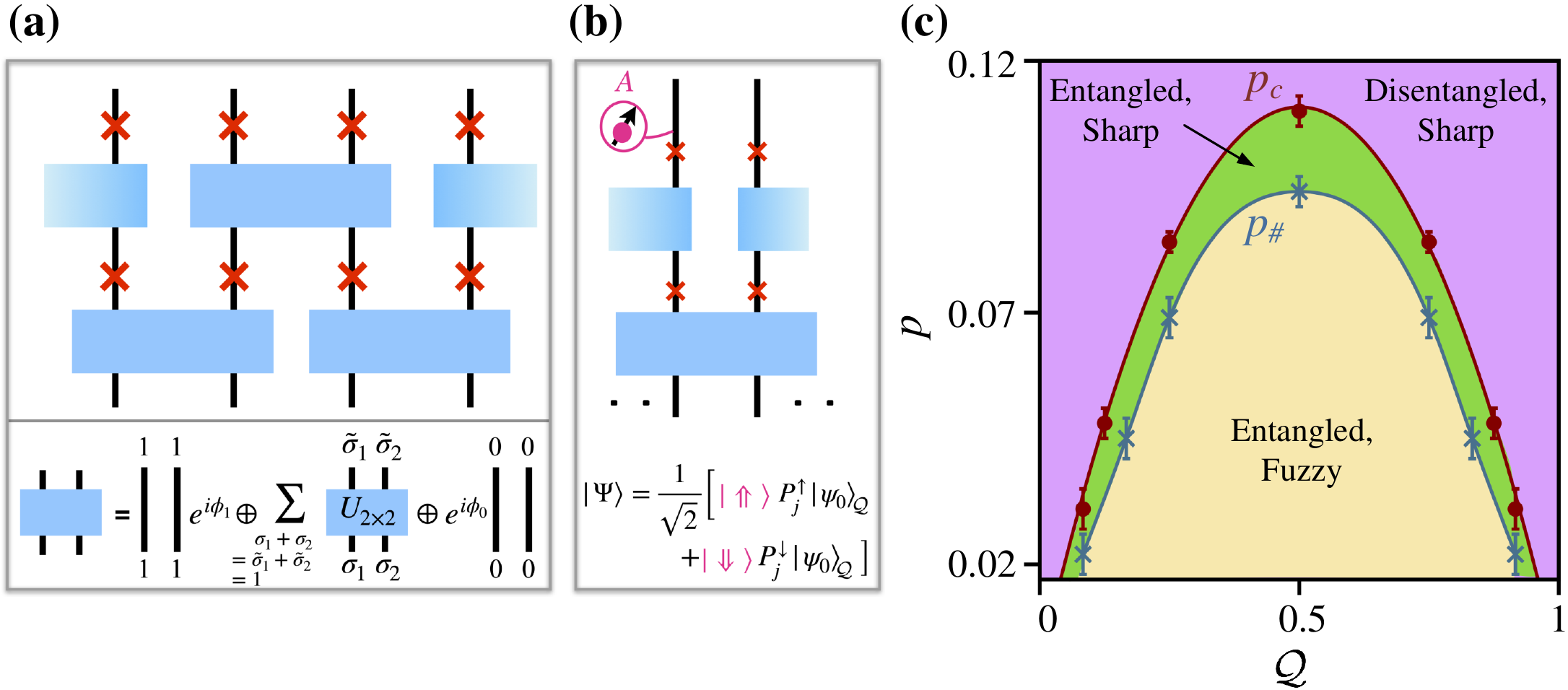}
  \caption{{\bf Model, ancilla qubit, and sector resolved phase diagram}. (a) shows brick-layer structure of the monitored quantum circuit consisting of U(1) symmetric Haar random gates (the blue bricks) and projective $\sigma^z$ measurements (the red crosses) at a rate $p$ on the 1D chain of qubits. The U(1) symmetric gate given in Eq.\ref{Ugate} acting on two neighbouring qubits ($\sigma_1$ and $\sigma_2$)is also shown. 
  (b) An ancilla ($A$) qubit is maximally entangled to two orthogonal states, $|\psi_0;j=\uparrow \rangle_{\mathcal{Q}} = P_j^{\uparrow} |\psi_0 \rangle_{\mathcal{Q}}$ and $|\psi_0;j=\downarrow \rangle_{\mathcal{Q}} = P_j^{\downarrow} |\psi_0 \rangle_{\mathcal{Q}}$ by projecting a single site $j$ of the state $|\psi_0\rangle_{\mathcal{Q}}$ to up and down state respectively. These states belong to same global charge density sector $\mathcal{Q}$ and the entanglement entropy of $A$ serves as the order parameter of the entanglement transition within a sector $\mathcal{Q}$.
  (c) shows the critical measurement rates, $p_c$ (red circles) for the entanglement transition and $p_{\#}$ (blue cross) for the charge-sharpening transition, for different charge densities $\mathcal{Q}$; $p_{\#}<p_c$ for all charge densities $\mathcal{Q}$. $p_c$ is obtained from the finite-size scaling of the ancilla entanglement entropy when $A$ is coupled in one charge sector while for $p_{\#}$, $A$ is coupled across sectors ($\mathcal{Q}$ and $\mathcal{Q}+1/L$). We fit $p_c(\mathcal{Q})$ to obtain $p_c(\mathcal Q)=0.44\mathcal Q(1-\mathcal Q)$ shown by red solid line while the blue solid line is simply a guide to eye. 
  }
   \label{fig:method}
  \end{figure*}

Exploring the properties of far-from-equilibrium, non-unitary dynamics provides a theoretical challenge that links the theory of open quantum systems and quantum information while also serving as a crucial bridge to understanding the limits of quantum information processing, including quantum error correcting codes.
Moreover, recently developed noisy intermediate scale quantum (NISQ) devices \cite{Preskill2018} can directly probe theoretical predictions regarding the nature and universality of quantum dynamic protocols.
In fact, NISQ devices are poised to probe the interplay of entangling unitary dynamics and disentangling projective measurements \cite{GoogleNISQ,monroeMIPT,google2023quantum,Expt_measurement} thanks to technical developments allowing for the possibility of local, mid-circuit measurements \cite{midcircuitMIPT}.
Exemplars of this paradigm are measurement-induced phase transitions (MIPTs) where the rate of mid-circuit measurements competing with the rate of entangling unitary dynamics (embodied through the quantum circuit model in Fig.~\ref{fig:method}(a)) drives a transition in the dynamics of the system's entanglement \cite{NahumPRX,FisherMIPTPRB,FisherMIPTPRB2,AltmanPRLMIPT,SmithPRBMIPT,GullansPRX,JianMIPT,fisherreview,RIslamSimulation,AliPRL}.
These transitions are analytically tractable in special limits \cite{NahumPRX,JianMIPT,AltmanPRLMIPT,VasseurAnalytic}, but numerical evidence for the generic dynamics involving qubits indicate a distinguished universality class characterized by a multifractal logarithmic conformal field theory \cite{Aidan_PRL,cardy,multifractalLudwig,LUDWIGCardyCFT,JianMIPT,AltmanPRB,VasseurAnalytic,FisherPRBCFT}.

A class of MIPTs were described in Ref.~\cite{GullansPRL} as ``purification transitions,'' (for many cases, these coincide with the aforementioned entanglement transitions).
Information theoretically, the ``mixed phase'' can robustly encode quantum information (which could be decoded) while the ``pure phase'' destroys any encoded information (i.e., it is theoretically impossible to decode).
Investigating the transition then not only tells you that the entanglement obeys a universal description, but also how quantum information is being destroyed.

This aspect can change when the dynamics are enriched with a symmetry. Recent studies have explored the effects of the presence of discrete, abelian, and non-abelian symmetry constraints in the hybrid dynamics~\cite{z2sym,Fermionic_Z2,PRX_U1,SU2,nonCircuit1,nonCircuitDiehl,nonCircuitDiehlPRX,nonCircuitMirlin}, 
resulting in an enriched  phase diagram, with distinct phases within volume-law and area-law entangled phases. In particular, with a continuous global U(1) (i.e., charge conserving) symmetry in a random monitored circuit ~\cite{PRX_U1,PRL_U1,PRB_U1}, a new transition can appear where the system can undergo ``charge sharpening'' at a measurement rate $p_\#$ slower than the rate at which it purifies $p_c$. 
Namely, an initial state that has equal weight in all charge sectors will collapse into a single sector on short (i.e., an $\sim O(1)$) time scale for $p>p_\#$ while still being volume law entangled. In contrast, in the charge fuzzy phase this collapse to a single sector occurs on an $\sim O(L)$ time scale for a linear system size $L$. 
The idea of ``sharpening'' is also not 
a property of the initial state
but is in fact a bulk phenomena which manifests itself in the dynamics of local operators \cite{PRL_U1,PRB_U1}.
Additionally, understanding the fate of charge-sharpening transition in non-circuit models remains an open interesting question~\cite{nonCircuit1,nonCircuitDiehl,nonCircuitDiehlPRX,nonCircuitMirlin}. 

Our present work addresses how a global $U(1)$ symmetry in monitored quantum circuits enriches the phase diagram in different symmetry sectors and affects the critical properties across the enriched phase diagram. In these dynamics, the charge sector that is labeled by charge density $\mathcal Q = Q/L$, where $Q =\sum_i (\sigma^z_i+1)$ is the total conserved charge of the chain of qubits where $\sigma_i^z$ is the $z$ Pauli matrix, is a (classical) bit of information that can be lost prior to any quantum information.
For the qubit system and $\mathcal{Q}=1/2$, these transitions appear to be close to one another (Ref.~\cite{PRX_U1} estimates $p_\# \approx 0.094(4) $ and $p_c\approx 0.110(3)$), and therefore any NISQ device which probes the loss of this quantum information at the entanglement transition $p_c$ 
may also probe some of the features of the nearby sharpening transition as the available system sizes may not be larger than the two correlation lengths ($\xi$ and $\xi_\#$) of each  respective transition.
Elucidating and understanding the critical properties of the MIPT  that one can discern in this setup is one of the major achievements of our present work.

The generic MIPT enriched by a global U(1) symmetry and resolved into individual charge sectors $\mathcal Q$ is shown in Fig.~\ref{fig:method} (c). 
Here, we say $p<p_\#(\mathcal Q)$ is defined as \emph{charge fuzzy} while $p>p_\# (\mathcal Q)$ is \emph{charge sharp} and
the transition occurs within the volume-law entangled phase of the system.
At larger measurement rates $p>p_c(\mathcal Q)$ the model undergoes an MIPT to an area-law entangled phase. Thanks to the perspective of a purification transition, we can putatively probe the  sharpening transition and the MIPT separately. In particular, by coupling an auxiliary ancilla qubit into the system that either couples across charge sectors (e.g. between $\mathcal Q$ and $\mathcal{Q} +1$)
 or within a given sector (e.g. between two distinct orthogonal spin states with equivalent $\mathcal{Q}$) as depicted in Fig~\ref{fig:method}(b), we should in principle be able to examine the sharpening transition and the entanglement transition separately.

However, the nature of the critical properties of the sharpening transition may have a significant impact on these conclusions. In particular, in Ref.~\cite{PRL_U1} it was shown that a modified version of these dynamics (in the limit of an infinite onsite Hilbert space dimension) has a charge sharpening transition which falls into the Berezinskii–Kosterlitz–Thouless (BKT)\cite{BKT_B,BKT_KT} universality class but has no effect on the entanglement transition which remains controlled by percolation. There are several important implications of this field theoretic prediction. Firstly, the critical nature of the BKT transition, makes the entire charge fuzzy phase ``critical'' in a similar fashion that terminates at the BKT transition~\cite{PRL_U1}. Second, this critical dynamics could in principle be seen in a single sector at late times, as probed through correlation functions (not entanglement measures) averaged over measurement outcomes~\cite{PRL_U1}.
Third, the finite size cross-over regime of the critical fan, defined by the correlation length ($\xi_\#$)  for a BKT transition is known to be broader than a second order transition ($\xi$), see Fig.~\ref{fig:BKT_illustration}. As a result, small system size calculations probing the MIPT could receive a contribution from the sharpening critical point (and vice versa).

In this manuscript, we present a sector-resolved analysis of the universality class of the entanglement transition  in a U(1) symmetry preserving random-Haar circuit with projective measurements (shown in Fig.~\ref{fig:method}(a)). In addition, we  present a detailed study of the MIPT and the critical sharpening properties in the middle sector. We adopted an efficient method to couple an ancilla to the charge-conserving circuit (shown in Fig.~\ref{fig:method}(b)). This allows us to study the critical properties in different charge sectors, even those which are far off from half-filling, allowing our finite-size numerics to access larger system sizes thanks to the constrained many-body Hilbert space. The main findings of our work are summarized below:
\begin{enumerate}
\item We compute the sector-resolved phase diagram in Fig. \ref{fig:method}(c) which shows that the system undergoes a charge-sharpening transition at the measurement rate $p_{\#}(\mathcal Q)$ followed by the volume-law to area-law entanglement transition at $p_c(\mathcal Q) > p_{\#}(\mathcal Q)$ in all charge-sectors. We find the functional form of $p_{c}(\mathcal Q) \propto \mathcal{Q}(1-\mathcal{Q})$. Each sector is found to have a Lorentz invariant critical point at both the entanglement and charge sharpening transitions. 
\item We present numerical results showing that the estimated critical exponents of the entanglement transition are anomalously large in the U(1) symmetric circuit compared to those without any symmetry~\cite{AidanPRB,Aidan_PRL} or percolation. To show this, we compute the anomalous scaling dimension exponent, $\eta$ using the mutual information of a pair of ancillas in section \ref{sec:corrl}. We also study the non-unitary log-CFT governing the entanglement transition in section \ref{sec:cft} and obtain the effective central charge ($c_{\rm{eff}}$) of the log-CFT from the free energy, the typical scaling dimension $x_1^{\rm{typ}}=\eta/2$ (finding excellent agreement with the ancilla computation) and the higher cumulants (e.g., the $2$nd cumulant $x_1^{(2)}$) that probes the multi-fractal nature of the correlation functions. These critical exponents are listed in table \ref{tab:exponents}. 
As the BKT transition is not expected to modify the multifractal scaling, the large difference the symmetry has on the estimate of $x_1^{(2)}$ and the multifractal spectrum (in Sec.~\ref{sec:cft}) will serve as our strongest evidence of the universality class of the MIPT being modified by the U(1) symmetry.

\begingroup
\squeezetable
\begin{table}
\begin{center}
\begin{ruledtabular}
\begin{tabular}{cccc}
~ & $U(1) $  & No-symmetry & BKT   \\
~ & symmetry & ($H$) & ($\#$)  \\
\hline
$c_{\rm{eff}}$ & 1.27(1) & 0.25(3) & 1  \\
$x_1^{\mathrm{typ}}=\eta/2$  & 0.28(2)  & 0.14(2) &  0.125 \\
$x_1^{(2)}$ & 0.65(2) & 0.15(2) & NA 
\end{tabular}
\end{ruledtabular}
\end{center}
\caption{A comparison of the critical exponents governing the entanglement transition with and without U(1) symmetry is shown. No-symmetry values taken from Ref.~\cite{Aidan_PRL}. Their known values in the BKT universality class are also listed. 
The large values of the exponents in presence of $U(1)$ symmetry compared to the usual Haar MIPT suggests that the symmetry constraint changes the universality class. We provide two possible interpretations of these results in Sec.~\ref{sec:illustration}.}
\label{tab:exponents}
\end{table}
\endgroup

\begin{figure*}[t]
\begin{center}
  \includegraphics[width=0.99\textwidth]{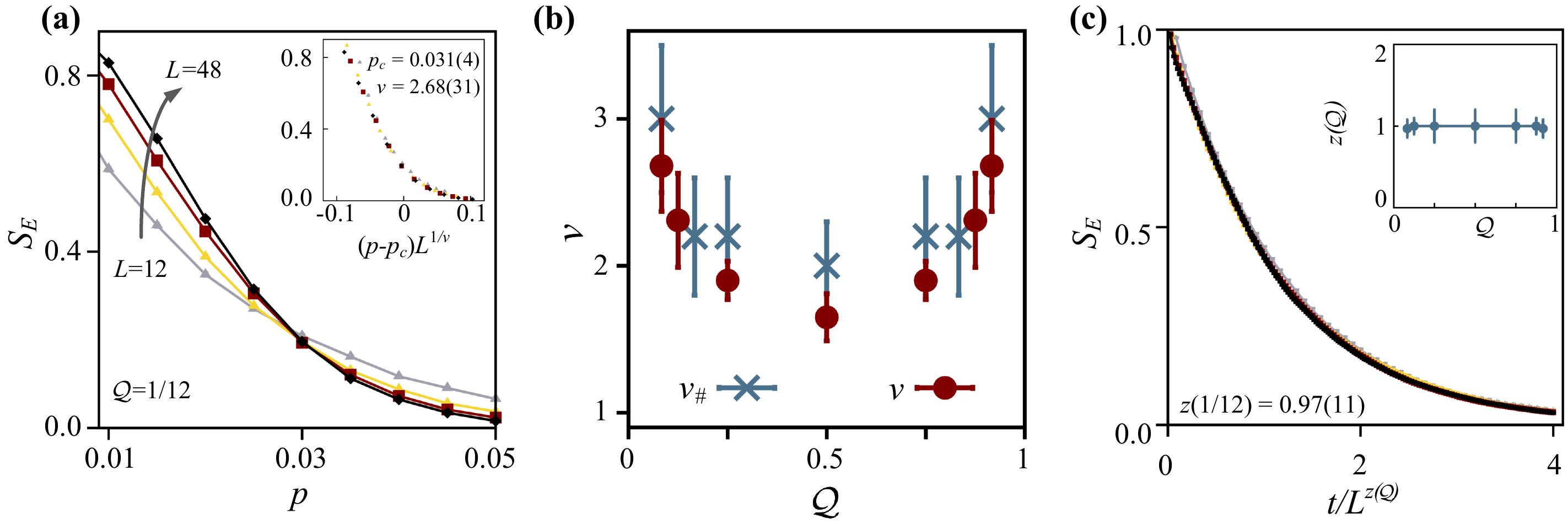}
  \caption{{\bf Entanglement Transition}. Entanglement entropy $S_{E}$ of the ancilla serves as the order parameter of the MIPT with $p$. (a) shows the crossing of $S_{E}$ vs $p$ for different $L=12$ to $48$ with total charge $Q=L/12$ taken at time $t=2L$. This yield $p_c=0.031(4)$ and $\nu=2.68(31)$ from the finite-size scaling collapse shown in the inset.
  (b) shows the correlation of the critical expo nents $\nu$ and $\nu_{\#}$ at $p_c$ (red circles) and $p_{\#}$ (blue cross) respectively; this data was obtained from the finite-size scaling of $S_E$ and $S_{\#}$ (shown in appendix \ref{secSM:OP} and \ref{secSM:BKT})  for different $\mathcal{Q}$ following Eqs.~\eqref{eq:SE} and \eqref{eq:Ssharp} respectively. 
  (c) shows the collapse of $S_{E}$ vs $t/L^{z(\mathcal{Q})}$ for different $L=12$ to $48$ at $\mathcal{Q}=1/12$. This gives $z(\mathcal{Q}=1/12)=0.97$. $z \approx 1$ for all sectors as shown in the inset demonstrating each transition is Lorentz invariant. }
  
   \label{fig:ancilla}
   \end{center}
  \end{figure*}

\item By studying correlation functions of the conserved charge, we show that critical sharpening fluctuations are present within a single sector and provide evidence that the nature of the sharpening transition is BKT-like with the expected quantized jump in the stiffness at the transition in section \ref{subsec:BKT}.  Moreover, they remain critical at the MIPT within the accessible system sizes available. This is consistent with a bulk (i.e. independent of the initial state) description of charge sharpening \cite{PRL_U1,PRB_U1}.

\end{enumerate}

This leads us to provide two different perspectives on the putative coexistence of the MIPT and charge sharpening transitions in Sec.\ref{sec:illustration}: (a) one possible scenario is that these two transitions are separate in the thermodynamical limit. But at finite length scale, the critical measurement rates of MIPT ($p_c$) and charge-sharpening transition $p_{\#}$ are very close by and the above critical exponents ($x_1^{(\rm{typ})}$ and $c_{\rm{eff}}$) we probed here through finite size scaling show a combination of both  criticalities, due to a broad finite-size critical fan as one would expect from a BKT transition (see Fig.~\ref{fig:BKT_illustration}). 

Within this interpretation,
these values are offset by the nearby sharpening transition, which by assuming it is a BKT transition allows us to obtain similar in magnitude (within our numerical accuracy) critical properties to the MIPT without any symmetry for $x_1^{\mathrm{typ}}$ and $c_{\mathrm{eff}}$. Contrarily, the critical exponent ($x_1^{(2)}$) signalling the multi-fractality aspect of correlation spectrum is expected to be unaffected by the nearby BKT transition (as it is known to  not be a multifractal CFT) and we observed large multi-fractality in the $U(1)$ symmetric dynamics compared to the non-symmetric counterpart. Hence, thanks to the multifractal nature of the MIPT and the BKT universality at the sharpening transition, we are able to firmly provide strong numerical evidence that the MIPT in the presence of a U(1) symmetry belongs to its own distinct universality class independent of the two interpretations we provide. (b)  A second possible scenario is that MIPT and charge-sharpening transition occur at the same measurement rate. In this scenario, all the exponents are distinct in in $U(1)$ symmetric dynamics compared to the usual Haar MIPT. This again ascertains our previous conclusion that the presence of $U(1)$ symmetry changes the universality class from the usual Haar MIPT. However, at present this second scenario is a less likely one as it is in contradiction with 
 results obtained in Refs.~\cite{PRX_U1,PRL_U1} in a suitably defined large onsite Hilbert space ($d\rightarrow \infty$ ) limit, which finds that the entanglement transition and charge-sharpening transition are separate in this limit.

\section{Model and Ancilla Probes}

 The quantum circuit consists of a 1D chain of qubits with local charge $q_i=(\sigma_i^z +1)/2$. The time-evolution of the circuit repeats the brick-layer structure, shown in Fig.~\ref{fig:method}(a), consisting of (i) the entangling unitary gates $U_{i,i+1}$ acting on the bond of the nearest-neighbor sites $i$ and $i+1$. $U_{i,i+1}$ conserve the total charge $q \in 0,1,2$ on the bond and are chosen from the set of generic Haar-random unitary gates of the form
 \begin{eqnarray}
\setlength\arraycolsep{0pt}
U = \begin{pmatrix}
e^{i\phi_0} &      0    &   0    \\
       0     & \boxed{U_{2 \times 2}} &    0   \\
     0       &     0  &     e^{i\phi_1}    \\     
\end{pmatrix}
\label{Ugate}
 \end{eqnarray}

where $U_{2 \times 2}$ is a $2 \times 2$ Haar-random matrix and $\phi_0$ and $\phi_1$ are random phases.
(ii) each site is subjected to projective $\sigma_i^z $ measurement with a rate $p$. The non-unitary dynamics conserve the global U(1) charge $Q=\sum_{i=1}^{L}q_i$. This constrains the dimension of the effective Hilbert space accessible during the dynamics if the initial condition is chosen from a fixed $\mathcal{Q}=Q/L$ sector. This 
allows us to explore larger finite system sizes for the sectors away from half-filling. We use periodic boundary conditions unless mentioned otherwise.

\subsection{Ancilla to study the entanglement transition} \label{subsec:singleanc}
In our present work, we aim to extract critical exponents that correspond to a power law decay of the correlation functions of the CFT, via mutual information between two ancilla qubits. To couple an ancilla to the U(1) symmetric circuit in a fixed charge sector, previous work in Ref.~\cite{PRX_U1} utilized a ``bond-ancilla" method where a single ancilla is coupled to a ``bond'' between two neighboring qubits that requires post-selecting to couple into  only the $\lvert\uparrow\downarrow\rangle$ and $\lvert\downarrow\uparrow\rangle$ states. Extending this approach to two ancillas as we need to compute their mutual information suffers from low statistics due to post-selecting now on two bonds.

To circumvent this issue, we implement an approach that utilizes a ``single-site ancilla'' protocol schematically shown in Fig.~\ref{fig:method}(b) by entangling each ancilla qubit to a single site of the system, yet conserving the global charge. We initialize the system in a random Haar state $\lvert\psi_0 \rangle_{{Q}}$ in the global charge sector ${Q}$ (or charge density sector $\mathcal{Q}$). We then create two orthogonal states $\lvert\psi_0;j=\uparrow \rangle_{{Q}} = P_j^{\uparrow} \lvert\psi_0 \rangle_{{Q}}$ and $\lvert\psi_0;j=\downarrow \rangle_{{Q}} = P_j^{\downarrow} \lvert\psi_0 \rangle_{{Q}}$ by projecting a single site $j$ of the system to up and down state respectively. We note that the states $\lvert\psi_0;j=\uparrow \rangle_{{Q}}$ and $\lvert\psi_0;j=\downarrow \rangle_{{Q}}$ remain in the same global charge sector ${Q}$. The ancilla qubit with two orthogonal states $\lvert\Uparrow \rangle $ and $\lvert\Downarrow\rangle$ is now maximally entangled to the site $j$ as,
\begin{equation}
 \lvert\Psi \rangle = \frac{1}{\sqrt{2}}\Big[ \frac{\lvert\psi_0;j=\uparrow \rangle_{{Q}}}{\| \lvert \psi_0;j=\uparrow\rangle \|} \lvert\Downarrow\rangle +  \frac{ \lvert\psi_0;j=\downarrow \rangle_{{Q}}}{\| \lvert \psi_0;j=\downarrow  \rangle \|}  \lvert\Uparrow\rangle\Big].
 \label{eq:ancilla_MIPT}
\end{equation}
This single-site ancilla method allows us to get around the post-selection problem as we increase the system size and hence can be used to probe the entanglement transition at different global conserved charge sectors, including those which are away from the middle sector ($Q=L/2$ or $\mathcal Q=1/2$).

\begin{figure*}[t]
  \includegraphics[width=0.99\textwidth]{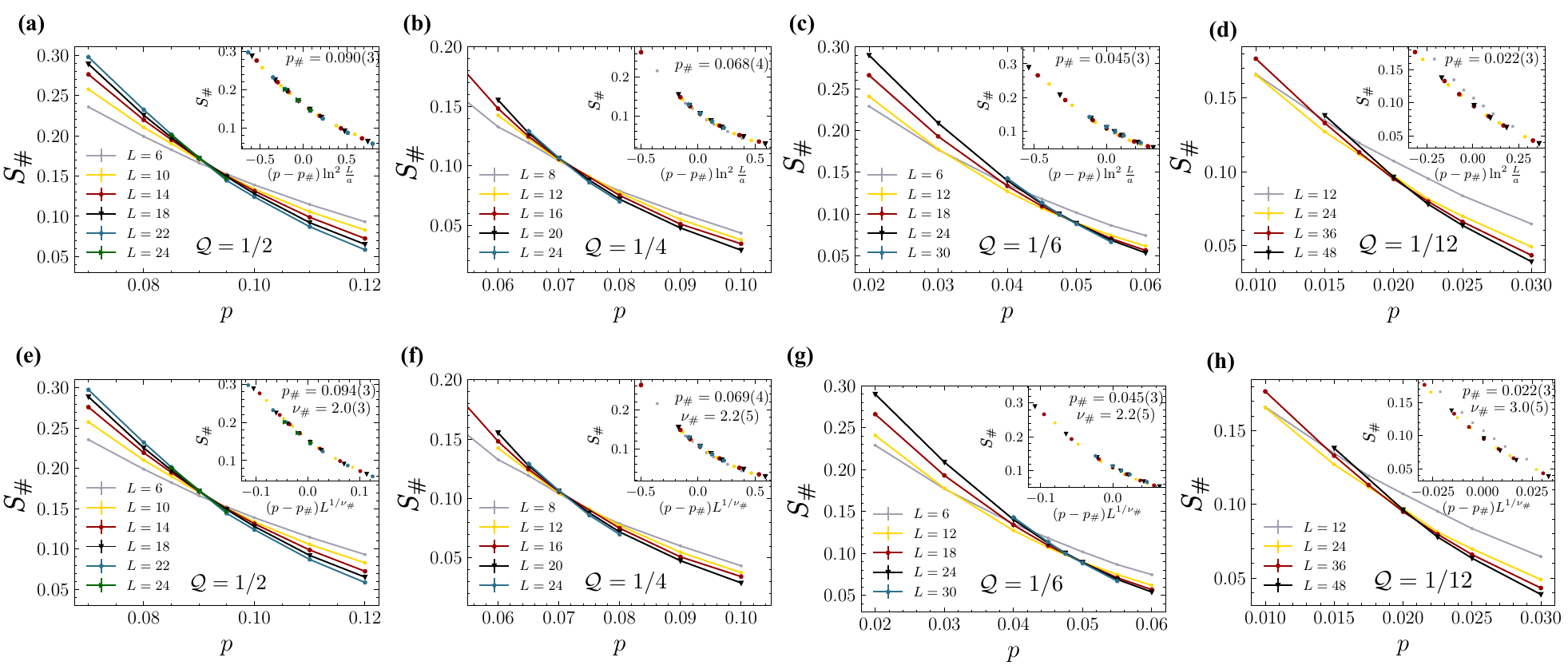}
  \caption{Charge-sharpening order parameter $S_\#$  vs $p$. The panels (a), (b), (c), (d) are collapsed with the BKT scaling ansatz Eq.\ref{eq:SharpLog}  (more precisely Eq.~\ref{eq:BKT_scaling}) at $\mathcal{Q}=1/2, 1/4, 1/6$ and $1/12$, respectively. The panels (e), (b), (c), (d) are collapsed with the second-order phase transition ansatz Eq.\ref{eq:Ssharp} (more precisely Eq.~\ref{eq:scaleKun}) at $\mathcal{Q}=1/2, 1/4, 1/6$ and $1/12$, respectively. The fitted $p_c(\mathcal{Q})$ and $\nu(\mathcal{Q})$ shown in the insets.
  }
   \label{fig:charge_sharpening_collapse}
  \end{figure*}  

\subsection{Ancilla to study the charge-sharpening transition} \label{subsec:anc_CS}
To probe the charge-sharpening (CS) transition at a given charge sector $Q$, we couple the ancilla to two neighboring charge sectors, ${Q}$ and ${Q}+1$ (or charge density sectors $\mathcal{Q}$ and $\mathcal{Q}+1/L$) \cite{PRX_U1},
\begin{equation}
 \lvert\Psi \rangle_{\#} = \frac{1}{\sqrt{2}}\left[ \frac{\lvert\psi_0\rangle_{{Q}}}{\| \lvert \psi_0\rangle_{{Q}} \|} \lvert\Uparrow\rangle +  \frac{ \lvert\psi_0\rangle_{{Q}+1}}{\| \lvert \psi_0\rangle_{{Q}+1} \|}  \lvert\Downarrow\rangle\right].
 \label{eq:ancilla_sharp}
\end{equation}
Here $|\psi_0\rangle_{{Q}}$ and $|\psi_0\rangle_{{Q}+1}$ are two initial states belonging to two different charge sectors ${Q}$ and ${Q}+1$ which maximally couple to two orthogonal states of the ancilla $|\Uparrow\rangle $ and $|\Downarrow\rangle $ respectively.

\section{Phase Diagram:} \label{sec:phases}
The  model introduced in the previous section exhibits a phase diagram~\cite{PRX_U1,PRB_U1,PRL_U1} that depends on the conserving charge sector as we show in Fig.~\ref{fig:method}(c). Focusing on a fixed charge sector or coherent superpositions over all charge sectors, with increasing $p$ the model exhibits two distinct phase transitions: an entanglement phase transition from the volume-law to area-law phase at the critical measurement rate $p_c$ which is preceded  by a charge sharpening transition at the critical measurement rate $p_{\#}< p_c$ in the volume-law phase.

\subsection{Entanglement Transition}
In this sub-section, we couple an ancilla qubit to the circuit (at $t=0$) initialized in a fixed charge density sector $\mathcal{Q}$ following the protocol described in Eq.\ref{eq:ancilla_MIPT}. We calculate the time-evolution of the entanglement entropy $S_{E}(t,L;\mathcal Q)$ of the ancilla qubit with the system. $S_{E}(t,L;\mathcal Q)$ follows a finite size scaling (FSS) ansatz near the entanglement transition point $p=p_c(\mathcal{Q})$,
\begin{equation}
    S_{E}(t,L;\mathcal Q)\sim h_{\mathcal Q}[(p-p_c(\mathcal{Q}))L^{1/\nu(\mathcal Q)},t/L^{z(\mathcal Q)}]
    \label{eq:SE}
\end{equation}
where $h_{\mathcal{Q}}(x,y)$ is a two-parameter universal scaling function with a correlation length exponent $\nu(\mathcal Q)$ and dynamic exponent $z(\mathcal Q)$ for different $\mathcal{Q}$ sectors. $\nu(\mathcal Q)$ governs the divergence of the correlation length $\xi \sim (p-p_c)^{-\nu} $ at the entanglement transition. While finding $p_c(\mathcal{Q})$ and $\nu(\mathcal Q)$, we fix $t=2L$ and perform a one-parameter FSS analysis with $(p-p_c(\mathcal{Q}))L^{1/\nu(\mathcal Q)}$ and we then check its consistent through $t/L$ scaling at $p_c(\mathcal{Q})$.

In the steady state, $S_{E}$ serves as the order parameter of the entanglement transition. We study variation of $S_{E}$ with $p$ for different system sizes $L$ commensurate with the global charge density sector $\mathcal{Q}$. Following previous work \cite{PRX_U1}, we examine the $p$ dependence of $S_{E}$ by fixing the aspect ratio of time and length of the system ($t/L=2$) and ensure our crossing and collapse in FSS analysis is unaffected by this choice.  
At the middle sector ($\mathcal Q = 1/2)$, our scaling analysis yields $p_c=0.110(3)$ and $\nu=1.65(16)$ (see appendix \ref{secSM:OP}) which are in  agreement with the ``bond-ancilla" method results \cite{PRX_U1} and thus establishes the consistency of the ``single-site ancilla" method in the conserving circuit.
In Fig.~\ref{fig:ancilla}(a) we show $S_{E}$ vs $p$ at $Q=L/12$ for $L=12$ to $48$ and the inset shows the scaling collapse with $p_c=0.031(4)$ and $\nu=2.68(31)$. We summarize the dependence of the global conserved charge density $\mathcal Q=Q/L$ on the entanglement critical point $p_c(\mathcal Q)$ in Fig.~\ref{fig:method}(c) (with red circles) and the corresponding correlation length exponent $\nu(\mathcal Q)$ in Fig.~\ref{fig:ancilla}(b) (with red circles) respectively. Fig.~\ref{fig:method}(c) shows that changing the value of the conserved charge density $\mathcal Q$ from the half-filling condition weakens the entangling unitary gates and $p_c$ decreases. $p_c(\mathcal Q)$ fits well with the variance of the Binomial distribution $p_c(\mathcal Q)=0.44\mathcal Q(1-\mathcal Q)$ shown by red solid line.  
Fig.~\ref{fig:ancilla}(b) shows that $\nu$ increases as $\mathcal{Q}$ deviates from the half-filling, but with large statistical error at $\mathcal{Q}$ away from half-filling. 

To this end, at the critical point $p_c(\mathcal Q)$, we collapse $S_{E}(t)$ with $t/L^{z(\mathcal Q)}$ which gives an estimation of $z(\mathcal Q)$ for each charge sector. Fig.~\ref{fig:ancilla}(c) shows the collapsed data at $\mathcal{Q}=1/12$ for $L=12,24,36$ and $48$. Our numerical analysis finds $z(\mathcal Q) \approx 1$ for all $\mathcal{Q}$ sectors as shown in the inset of Fig.~\ref{fig:ancilla}(c). As Ref.\onlinecite{JianMIPT, Aidan_PRL} established the requirement of $z=1$ for conformal invariance, our numerical results suggests conformal invariance across all charge sectors we have considered.

\subsection{Sharpening Transition}

In this sub-section, to study the charge-sharpening transition at a charge sector $Q$ (or charge density sector $\mathcal{Q}$), we couple an ancilla qubit to the circuit (at $t=0$) in a maximally entangled state with two neighbouring charge sectors, ${Q}$ and ${Q}+1$ following the protocol described in Eq.\ref{eq:ancilla_sharp}. In the steady state, the entanglement entropy of this ancilla, $S_{\#}$ serves as the order-parameter of the charge-shrapening transition. $S_{\#}(t,L;\mathcal Q)$ follows a FSS ansatz near the charge-sharpening transition point $p=p_{\#}(\mathcal{Q})$,
\begin{equation}
    S_{\#}(t,L;\mathcal Q)\sim g_{\mathcal{Q}}[(p-p_{\#}(\mathcal{Q}))L^{1/\nu_{\#}(\mathcal{Q})},t/L^{z_{\#}(\mathcal Q)}]
    \label{eq:Ssharp}
\end{equation}
 where $g_{\mathcal{Q}}(x,y)$ is a two-parameter universal scaling function with a correlation length exponent $\nu_{\#}(\mathcal Q)$ and dynamic exponent $z_{\#}(\mathcal Q)$ for different $\mathcal{Q}$ sectors. 
 We study the $p$ dependence of $S_{\#}$ fixing $t=2L$ and perform a one-parameter scaling ansatz with $(p-p_{\#}(\mathcal{Q}))L^{1/\nu_{\#}(\mathcal{Q})}$.
 The dependence of $p_{\#}(\mathcal Q)$ and $\nu_{\#}(\mathcal Q)$ on global conserved charge density $\mathcal Q=Q/L$ are shown in Fig.~\ref{fig:method}(c) (with blue cross) and in Fig.~\ref{fig:ancilla}(b) (with blue cross) respectively (data collapse shown in Fig.\ref{fig:charge_sharpening_collapse}(e)-(h)). For all $\mathcal{Q}$ sectors, the charge-sharpening transition precedes the entanglement transition, $p_{\#}(\mathcal Q)<p_c(\mathcal Q)$.

Interestingly, for $\mathcal{Q}$ sectors away from half-filling, the two critical points are approaching each other within their  error bars. Moreover, our estimate of the critical exponents $\nu$ and $\nu_{\#}$ become strongly dependent on the sector and we cannot discern them at the small system sizes we can reach. As the error bars on $\nu$ and $\nu_{\#}$ are growing as we move away from the middle sector and the largest deviation from the middle sector to $\mathcal Q = 1/12$ differ by at most `two $\sigma$', we cannot rule out the possibility that these exponents do not depend on the sector and the trend we witness is a finite size effect. 

These results in conjunction with the recent prediction of the charge sharpening transition being a BKT transition in the large on-site Hilbert space limit, this large $\nu_{\#}$ at the small charge density sectors suggests that our scaling ansatz could be incorrect and instead is suggestive that the power law should be replaced with the logarithmic dependence due to the nature of the BKT correlation length. In Fig.\ref{fig:charge_sharpening_collapse}(a)-(d) we find that a FSS analysis with BKT scaling ansatz,
\begin{equation}
S_{\#}(L;\mathcal Q)\sim g_{\mathcal Q}[(p-p_{\#}(\mathcal Q))(\log L/a)^2],
\label{eq:SharpLog}
\end{equation}
also gives reasonably good collapse with an estimated value of $p_{\#}(\mathcal{Q})$ matching with the result from the second-order scaling anstaz (see Eq.\ref{eq:Ssharp}) within error-bar. At the presently available system sizes it is not possible for us to discern which scenario (second order vs BKT sharpening transition) is correct. Nonetheless,  in Sec.~\ref{subsec:BKT} we provide additional evidence of the BKT scenario through a finite size estimate of the predicted jump in the stiffness.

Independent of which scaling form Eq.~\eqref{eq:Ssharp} or \eqref{eq:SharpLog} is ultimately correct, we are able to probe the dynamical critical exponent of the sharpening transition.  In our study, we observe an emergent Lorentz invariance at the charge-sharpening phase transition with a generic $\mathcal{Q}$ sector. To demonstrate this, we study $S_\#(t)$ as a function of the rescaled time $t/L$ across various system sizes, keeping the $\mathcal{Q}$ sector constant. The effective collapse of $S_\#(t)$ signifies that $z=1$ is valid for typical $\mathcal{Q}$ sectors. For a detailed explanation, refer to appendix \ref{secSM:BKT}.

\section{Critical Properties of the Entanglement Transition}

\subsection{Mutual Information Probe of the Correlation Functions}\label{sec:corrl}

We use the ``single-site ancilla" method explained in the previous section to compute the correlation function at the critical measurement rate of the entanglement transition. We initialize the spin-chain in a fixed global charge density sector $\mathcal{Q}$ and evolve it under U(1) symmetric monitored dynamics up to late time $t_0=2L$. In the steady state, we couple two ancillas $A$ and $B$ to the circuit at $(r_1,t_1)$ and $(r_2,t_2)$ and calculate their mutual information given by, $I_n(A,B)= S_n(A)+S_n(B)-S_n(A\cup B)$, where $S_n$ denote the Renyi entropy of order $n$. To compute the order parameter correlation, we set $t_1=t_2=t_0$ and fix $|r_1-r_2|=b L$ with $b$ being a constant. The correlation function obeys a one-parameter finite-size scaling ansatz with time $t-t_0$ as, 
\begin{equation}
C(t-t_0,L/2,Q) \sim \frac{1}{L^{\eta(Q)}} k_Q\Bigg(\frac{t-t_0}{L}  \Bigg).
\label{eq:eta}
\end{equation}

\begin{figure}
\centering
   \includegraphics[width=0.95\columnwidth]{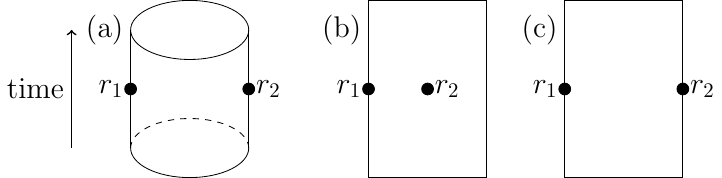}.
   \caption{Depiction of geometries for the pair of ancilla qubits used in the computation of $\eta$. Respectively, (a) $\eta$, (b) $\eta_\perp$, and (c)  $\eta_\parallel$.}
   \label{fig:eta_geometries}
\end{figure}

\begin{figure}[t!]
  \includegraphics[width=0.8\columnwidth]{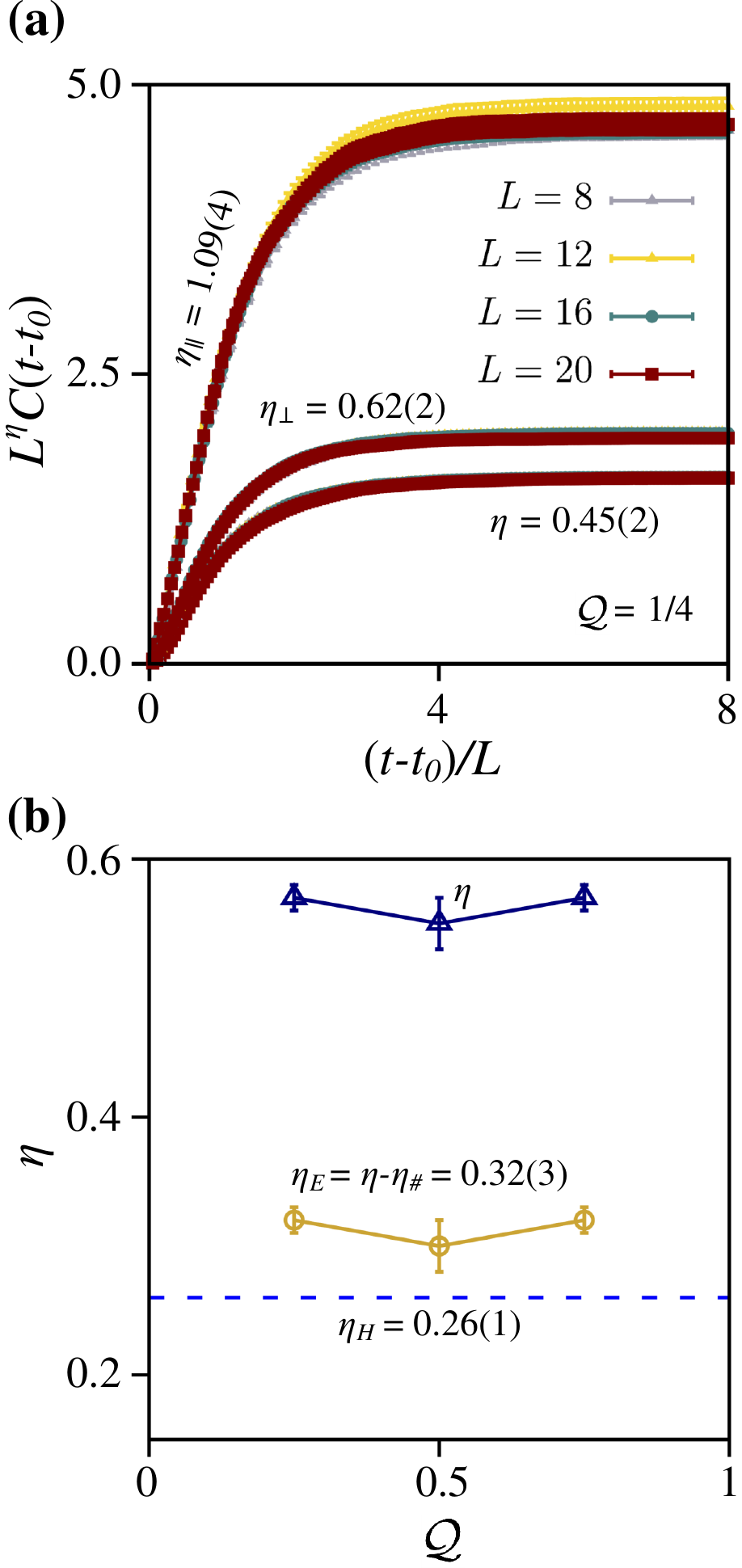}
  \caption{{\bf Probing $\eta$}. Mutual information $C(t-t_0)$ between two ancilla qubits coupled to the monitored U(1) symmetric circuit in the same charge sector $\mathcal{Q}=Q/L$ at $p=p_c(\mathcal{Q})$. (a) shows the scaling collapse of $C(t-t_0)$ following Eq.~\eqref{eq:eta} for different system sizes from $L=8$ to $20$ at $\mathcal{Q}=1/4$ for different geometries of the circuit: $\eta=0.45(2)$ for periodic BC at $|r_1-r_2|=L/2$, $\eta_{\parallel}=1.09(4)$ for open BC at $|r_1-r_2|=L$ and $\eta_{\perp}=0.62(2)$ for open BC at $|r_1-r_2|=L/2$. Geometries follow Fig.~\ref{fig:eta_geometries}, and we use the von Neumann entropy for this plot. 
  (b) shows the dependence of the bulk critical exponent $\eta$ on the charge sector with fixed density $\mathcal{Q}$ (at $n \rightarrow \infty$). $\eta$ exponent remains stable across all $\mathcal{Q}$ sectors by blue triangles. We also show the values $\eta_E$ once we subtract the known BKT values $\eta_{\#}=0.25$ by yellow circles. The blue dashed line show the value $\eta_H=0.26(1)$ in absence of symmetry in the circuit obtained from Ref.~\cite{AidanPRB}.} 
   \label{fig:eta}
   \end{figure}

Here, we calculate three different anomalous dimension exponent $\eta$ depending on the spatial separation and the boundary conditions (BC) in the circuit: (a) $\eta$ for $|r_1-r_2|= L/2$ with periodic boundary condition, (b) $\eta_{\perp}$ for $|r_1-r_2|= L/2$ with open boundary condition and (c) $\eta_{\parallel}$ for $|r_1-r_2|= L$ with open boundary conditions and geometries given in Fig.~\ref{fig:eta_geometries}.
The scaling collapse of these three cases is shown in Fig.~\ref{fig:eta}(a) at $Q=L/4$ for $L=8$ to $20$ for $n=1$. In Table~\ref{tab:eta} we show the dependence of the three anomalous dimension exponents $\eta,\eta_{\perp},\eta_{\parallel}$ on the Renyi index $n$. With increasing $n$, the three exponents saturate to the values (e.g., $\eta=0.57(1)$, $\eta_{\parallel}=1.27(4)$, $\eta_{\perp}=0.77(3)$ at $Q=L/4$). This shows that the finite-size estimation of $\eta, \eta_{\perp},\eta_{\parallel}$ in U(1) symmetric case are distinct from their corresponding value in the monitored Haar-random circuit without the U(1) symmetry constraint presented in Ref.~\onlinecite{AidanPRB}.

\begingroup
\squeezetable
\begin{table}
\begin{center}
\begin{adjustbox}{width=0.98\columnwidth,center}
\begin{tabular}{ccc|cc|cc}
\cline{1-7}
 n&\multicolumn{2}{c}{ $\eta $ } & \multicolumn{2}{c}{ $\eta_{\parallel}$} &  \multicolumn{2}{c}{ $\eta_{\perp}$}  \\
\cline{1-7}
 & $\mathcal{Q}=1/2$ &$1/4$  & $1/2$ & $1/4$  & $1/2$ & $1/4$  \\
\cline{2-3}
\cline{4-5}
\cline{6-7}
1&0.46(2)& 0.45(2) &0.97(2) & 1.09(4) &  0.61(1)&0.62(2) \\
2 & 0.55(3) & 0.54(1) & 1.13(3)&1.27(4) &0.72(3)& 0.71(3)  \\
5 &0.57(3) & 0.57(1) & 1.17(3)& 1.27(3) & 0.75(2) & 0.77(3) \\
$\infty$ &0.57(3)& 0.57(1) & 1.17(3) & 1.27(4) & 0.76(2)& 0.77(3) \\
\cline{2-3}
\cline{4-5}
\cline{6-7}
 No-symmetry & \multicolumn{2}{c} {0.26(1)} & \multicolumn{2}{c} { 0.49(2)} &  \multicolumn{2}{c} {0.34(1)} \\
 \cline{1-7}
\end{tabular}
\end{adjustbox}
\end{center}
\caption{We tabulate the anomalous scaling dimension exponents $\eta$ together with the surface critical exponents $\eta_{\parallel}$ and $\eta_{\perp}$ using open BC with $|r_1-r_2|=L$ and $|r_1-r_2|=L/2$ respectively for two values of $\mathcal{Q}=1/2$ and $1/4$. For Renyi indices $n>1$, all the exponents converge to values that are distinct from those without any symmetry in the circuit. 
}
\label{tab:eta}
\end{table}
\endgroup

In Fig.~\ref{fig:eta}(b) we show the variation of the exponent $\eta$,  with the global charge $\mathcal{Q}$. The blue solid line with triangles shows that the $\eta$ exponent (for $n=\infty$) remains same over different charge sectors (see appendix~\ref{secSM:eta} for more details) but is different from value $\eta_H=0.26(1)$ without the symmetry shown by blue dashed line.

\subsection{Probing the log-CFT} \label{sec:cft}
In this section, we will study the non-unitary CFT that governs the MIPT ($p=p_c(\mathcal{Q})$) in the dynamics of the circuit. We compute its critical properties ranging from the effective central charge,  typical anomalous scaling dimension, and the higher cumulants to study the multi-fractal nature of the correlation functions at the transition. Our numerical results indicate that U(1) symmetry changes the universality class, yielding a distinct multi-fractal scaling at the entanglement transition. 

\subsubsection{Free Energy }

The non-unitary time-evolution of a monitored random circuit can be described by the transfer-matrix method within the paradigm of statistical mechanics \cite{FisherPRBCFT,Aidan_PRL,JianMIPT}. In the quantum circuit, each trajectory is defined by a particular initial condition, a set of unitary gates and a set of measurement location, time and outcomes indexed by $\vec{m}$. Time-evolution of each trajectory is described by Krauss operators, $K_{\vec{m}}(t)=\prod_{t'=0}^{t} \hat{P}^{\vec{m}}_{t'} U_{t'}$ where $U_{t'}$ and and $\hat{P}^{\vec{m}}_{t'}$ denote a layer of unitary gates and projective measurement operators at time-step $t'$. Each trajectory at time $t$ is characterized by history of all measurement outcomes starting from $t=0$ upto $t$, denoted by $\vec{m}= \{ \dots,m_3,m_2,m_1\}$ (where $m_k=\pm 1$) associated with a conditional Born probability $p_{\vec{m}}$. We calculate $p_{\vec{m}}(t)$ by multiplying the Born probabilities of the individual measurement operations starting from $t=0$ upto time $t$. The non-unitary time-evolution is described by $\rho(t)= K_{\vec{m}}(t) \rho(t=0) K_{\vec{m}}^{\dagger}(t)/Z_{\vec{m}}$, where the partition function of the quantum trajectory $Z_{\vec{m}}(t)= p_{\vec{m}}(t)= Tr[ K_{\vec{m}}(t) \rho(t=0) K_{\vec{m}}^{\dagger}(t)] = \sum_i \exp(\lambda_i^{\vec{m}}t)$. 
At late time, the dynamics is governed by the leading Lyapunov exponents $\lambda_0^{\vec{m}},\lambda_1^{\vec{m}},\dots$ of the transfer matrix. We will calculate trajectory-averaged Lyapunov exponents $\lambda_i= \sum_{\vec{m}}p_{\vec{m}}  \lambda_i^{\vec{m}}$. $\lambda_0$ is related to the free energy of the statistical model, $F=-\lambda_0 t$ which can be calculated from the Shannon entropy of the measurement record, $F(t)=-\sum_{\vec{m}} p_{\vec{m}}(t) \ln p_{\vec{m}}(t)$. In this section, we will analyze the behaviour of $F(t)$ at $p=p_c(\mathcal{Q}=1/2)$ in the late time limit to eliminate the effects of the initial conditions~\cite{Aidan_PRL}. We will present two different cases where we either initialize the system in a state with unconstrained global charge or a state in a fixed charge sector. These two different cases show qualitatively different behaviour in the scaling of the free energy with $L$ as we explain below.

Using the conformal invariance at the entanglement transition point, the circuit dynamics can be mapped to $(1+1)d$ non-unitary conformal field theory characterized by a universal number called the effective central charge $c_{\rm{eff}}$. It can be related to the free-energy  density $f=F/A$ obtained from the leading Lyapunov exponent $\lambda_0$ as \cite{Aidan_PRL},
\begin{eqnarray}
\frac{F(L)}{A}=-\frac{\lambda_0}{\alpha L}= f(L\rightarrow \infty)- \pi \frac{c_{\rm{eff}}}{6 L^2}+\dots,
\label{eqn:F}
\end{eqnarray}
where $A=\alpha L t$ is the area of the cylinder of the CFT defined from a circuit of spatial and temporal length $L$ and $t$ respectively with a space-time anisotropy factor $\alpha$ to be computed numerically (see Appendix~\ref{sec:alpha} for details). We stress that $c_{\rm{eff}}$ is not related to the coefficient of the logarithmic scaling of the bipartite entanglement entropy with the system size, the latter of which is related to the scaling dimension of a boundary changing operator~\cite{AbhishekWeak}. Due to the conservation law, the nature of the initial state has a strong dependence on free energy. As $F=-\ln Z$ where $Z$ is the partition function of the statistical mechanics model \cite{Aidan_PRL} that involves a trace over all states, we begin by considering initial states that are equal superpositions over all charge sectors and discuss the effects of projecting into a given charge sector afterward. 
\begin{figure*}[t]
  \includegraphics[width=\textwidth]{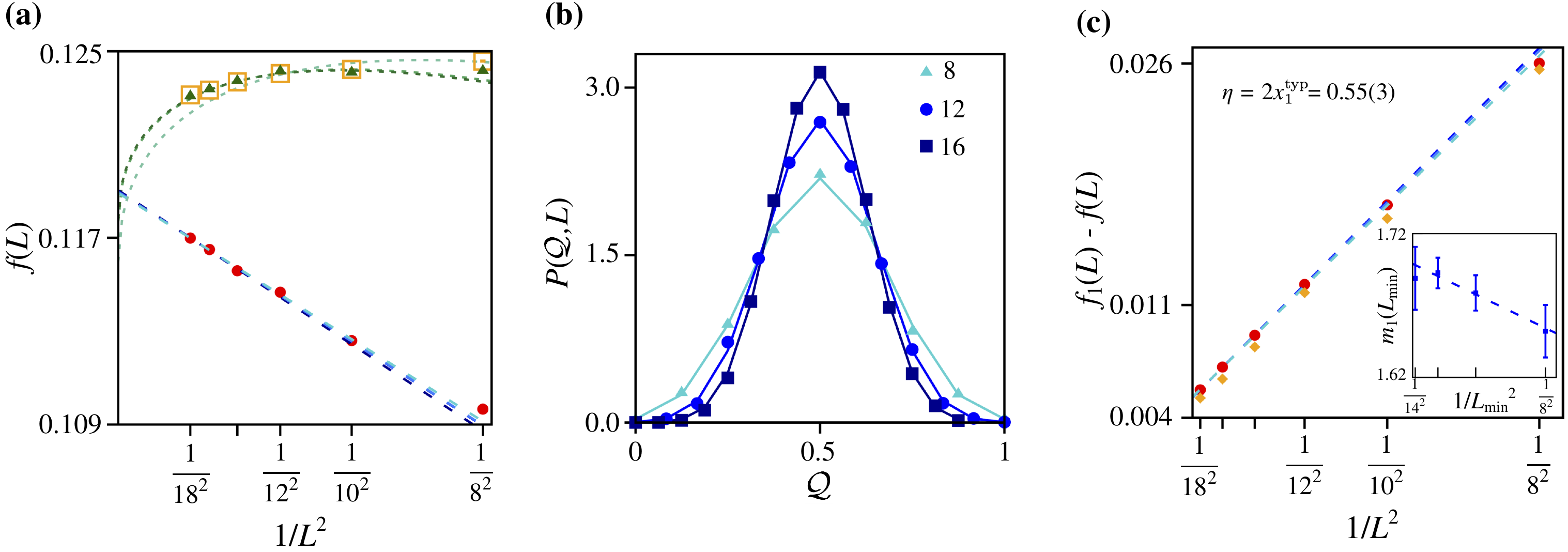}
  \caption{{\bf Scaling of the Free energy of the log-CFT}. At $p=p_c(\mathcal{Q}=1/2)=0.11$ of the monitored Haar-random circuit with U(1) symmetry, both (a) free energy density $f(L)$ and (c) difference between generalized free energies, $f_1(L)-f(L)$ show a $1/L^2$ scaling with system size. The slope of the former gives the effective central charge $c_{\rm{eff}}=1.27(1)$, while that of the latter gives the scaling dimension of the most relevant operator $x_1^{\rm{typ}}=2\eta=0.27(2)$ matching with that shown Fig.~\ref{fig:eta}. 
  (b) shows the rescaled probability distribution $P(\mathcal{Q},L)=L \times P(Q,L)$ vs $\mathcal{Q}$ where $P(Q,L)$ is the probability that an initial state with unrestricted global charge ends up in sector $\mathcal{Q}$ at late time ($t \gtrsim 4L$), which follows a binomial distribution dominated by the middle sector in $L\rightarrow \infty$ limit. The lines are not fits but simply plots of the rescaled binomial distribution.
  The initial conditions in (a) and (c) are randomly chosen from all $\mathcal{Q}$ sectors, except the orange squares in (a) and (c) respectively are the results selecting only those trajectories with $Q=L/2$ charge in the steady state. The green triangles in (a) correspond to the trajectories initialized and hence remained in $\mathcal{Q}=1/2$ sector throughout the evolution.
  } 
   \label{fig:CFT}
   \end{figure*}

 We first numerically compute the free energy density at $p=p_c(\mathcal{Q}=1/2)$ choosing the initial conditions (a product initial state or a Haar initial state) randomly with unconstrained total charge. As time evolves, we calculate free energy in each trajectory from the Born probabilities of the measurement events. Averaging over all such trajectories at a long time gives the total free energy $F(t)$ from all accessible trajectories evolved under the constrained dynamics. $F(t)$ grows linearly with time $t$ at late times with the slope $\lambda_0$ (we waited till time $t_o \sim 6L$ to reach the steady state and extract $\lambda_0$). The corresponding free energy density $f(L)$ decays as $1/L^2$ (consistent with the CFT prediction) as shown in Fig.~\ref{fig:CFT}(a) by red circles. We compute the slope of this curve $m_0(L_{\rm{min}})$ by systematically eliminating the smallest system sizes and considering data only from $L=L_{\rm{min}}$ to $L=18$. These fits are shown by dashed blue lines where the darker color corresponds to larger $L_{\rm{min}}$. This 
yields $m_0(L_{\rm{min}})= m_0(\infty)+b/L_{\rm{min}}^2$. The effective central charge is computed from $c_{\rm{eff}}=-6m_0(\infty)/\pi$. Our numerics predicts $c_{\rm{eff}}=1.27(1)$ which is significantly different from the result $c_{\rm{eff},H}=0.25(3)$ without U(1) symmetry in the circuit \cite{Aidan_PRL}.

We now bin the free energy into each charge sector $F^Q$ that it is projected into in the late time limit ($t\gg L$). We see that the probability that a given state ends up in a state with total charge $Q$ is precisely given by the binomial distribution $P(Q,L)=
\binom{L}{Q}
/2^L$. 
We plot it as $P(\mathcal{Q},L)=L \times P(Q,L)$ vs $\mathcal{Q}$ in Fig.~\ref{fig:CFT}(b).
Importantly, this shows that in the thermodynamic limit average quantities over initial states that consider all conserving sectors eventually converge to the middle sector. To relate the free energy over all sectors (i.e., the grand potential) and the free energy in the middle sector (i.e., the Gibbs free energy) requires introducing a chemical potential $\mu$, such that $F^\mathcal{Q}= F + \mu (Qt)$, where $Qt$ is the total charge in space-time. This implies that in addition to the $1/L^2$ dependence expected in Eq.~\eqref{eqn:F} there is a contribution from $\mu~\sim L^{-z}=1/L$ to all of the Lyapunov exponents in a fixed sector. Focusing on $f^{\mathcal Q}=F^Q/(\alpha Lt)$ for $\mathcal{Q}=1/2$ as shown in Fig.~\ref{fig:CFT}(a) open orange squares, we find that the data is well fit using the modified scaling form 
\begin{equation}
f^{\mathcal Q}=f^{\mathcal Q}(L\rightarrow \infty)+a\mathcal Q/ L-\pi \frac{c_{\rm{eff}}}{6L^2},
\label{eqn:Fmu}
\end{equation}
but with the same effective central charge, which provides evidence for our relation between the two free energies. We show this fit by green dashed lines for three different values of $L_{\rm{min}}= 8,10$ and $12$ where the darker color corresponds to larger $L_{\rm{min}}$, by fitting the data from $L=L_{\rm{min}}$ to $L=18$. 
Moreover, in the thermodynamic limit ($L \rightarrow \infty$), the Gibbs free energy ($f^{\mathcal Q}(L\rightarrow \infty)=0.118$) is approaching the free energy over all sectors ($f(L\rightarrow \infty)=0.119(1)$), allowing us to conclude that $f^{\mathcal Q}(L\rightarrow \infty)=f(L\rightarrow \infty)$ as expected. Lastly, the solid green triangles in Fig.~\ref{fig:CFT}(a) show the data when the system is initialized in $\mathcal{Q}=1/2$ sector and hence the circuit is restricted to $\mathcal{Q}=1/2$ throughout its evolution. These data points closely overlap with $f^{\mathcal{Q}=1/2}$ where the trajectories are binned to the middle sector at late times, establishing that the time-evolution is ergodic as we expect.

\subsubsection{Leading Scaling Dimension}

We now turn to probing the  scaling dimensions at the critical point. To do this, we compute the next leading Lyapunov exponent $\lambda_1$ which is related to the higher generalized free energy as $f_1=F_1/A=\lambda_1/\alpha L$. Following the procedure prescribed in Ref.\cite{Aidan_PRL} to compute $\lambda_1$ (in addition to $\lambda_0$), we create a pair of orthoganonal states $|v_0 \rangle$ and $|v_1 \rangle$. As time evolves, in each time-step of a trajectory, $|v_0 \rangle$ and $|v_1 \rangle$ are subjected to identical $U$ gates and measurements (locations and at which state it will be projected to), i.e identical Krauss operators $K_{\vec{m}}(t)$, governed sloely by the Born probabilities $p_{\vec{m}}$ of $|v_0 \rangle$. After each time-step, the pair of vectors need to be reorthogonalized with a Gram-Schmidt projector $P^{GS}_{t}$. The Born probability of the measurement records in the trajectory $\vec{m}$ corresponding the state $|v_1 \rangle$ is written as,
\begin{equation}
\tilde{p}_{\vec{m}}(t) = ||\Pi_{t'=0}^{t} P^{GS}_{t'} K_{\vec{m}}(t') |v_1 \rangle||^{2}.
\end{equation}
$F_1$ is calculated from the Born probabilities $\tilde{p}_{\vec{m}}$ of each measurement outcomes in the evolution of $|v_1 \rangle$, $F_1 (t)= \sum_{\vec{m}} \tilde{p}_{\vec{m}} (t) \ln \tilde{p}_{\vec{m}}(t)$.  

The difference between the two leading Lyapunov exponents grows as $1/L^2$,
\begin{equation}
    f_1(L)-f(L)=\frac{\lambda_1-\lambda_0}{\alpha L}= \frac{2\pi x_1^{\rm{typ}}}{L^2},
    \label{eq:xtypscale}
\end{equation}
shown in Fig.~\ref{fig:CFT}(c) by the red circles. 
This is used to calculate the scaling dimension of the most relevant operator $x_1^{\rm{typ}}$. 
We numerically estimate $x_1^{\rm{typ}}=0.27(2)$ (obtained from the slope $m_1(L_{min})$ shown in the inset by eliminating the smallest $L$s in the fit shown by dashed blue lines in Fig.~\ref{fig:CFT}(c)). $x_1^{\rm{typ}}$ is related to the decay of the bulk correlation function through the exponent $\eta=2 x_1^{\rm{typ}}=0.55(3)$. This gives an independent estimate of $\eta$ from CFT and importantly matches with our previous result from the order-parameter correlation function shown in Fig.~\ref{fig:eta}(b) (solid blue line with triangles).

Lastly, we considered probing the free energy differences after they are projected into the middle sector $f_1^{\mathcal{Q}=1/2}(L)-f(L)^{\mathcal{Q}=1/2}$ shown in Fig.~\ref{fig:CFT}(c) by orange squares and find that the chemical potential contribution, $\mu$ in Eq.~\eqref{eqn:Fmu}, cancels precisely leaving us with an additional estimate of $\eta=2x_1^{\rm{typ}}=0.51(4)$ (using the same fitting form given in Eq.~\eqref{eq:xtypscale}) that is in good agreement with that obtained from $f_1(L)-f(L)$ with $\eta=2x_1^{\rm{typ}}=0.55(3)$ sampled over all charge sectors (see appendix \ref{secSM:f1} for more details). 

\subsubsection{Multi-fractality}
\begin{figure}[t]
  \includegraphics[width=0.85\columnwidth]{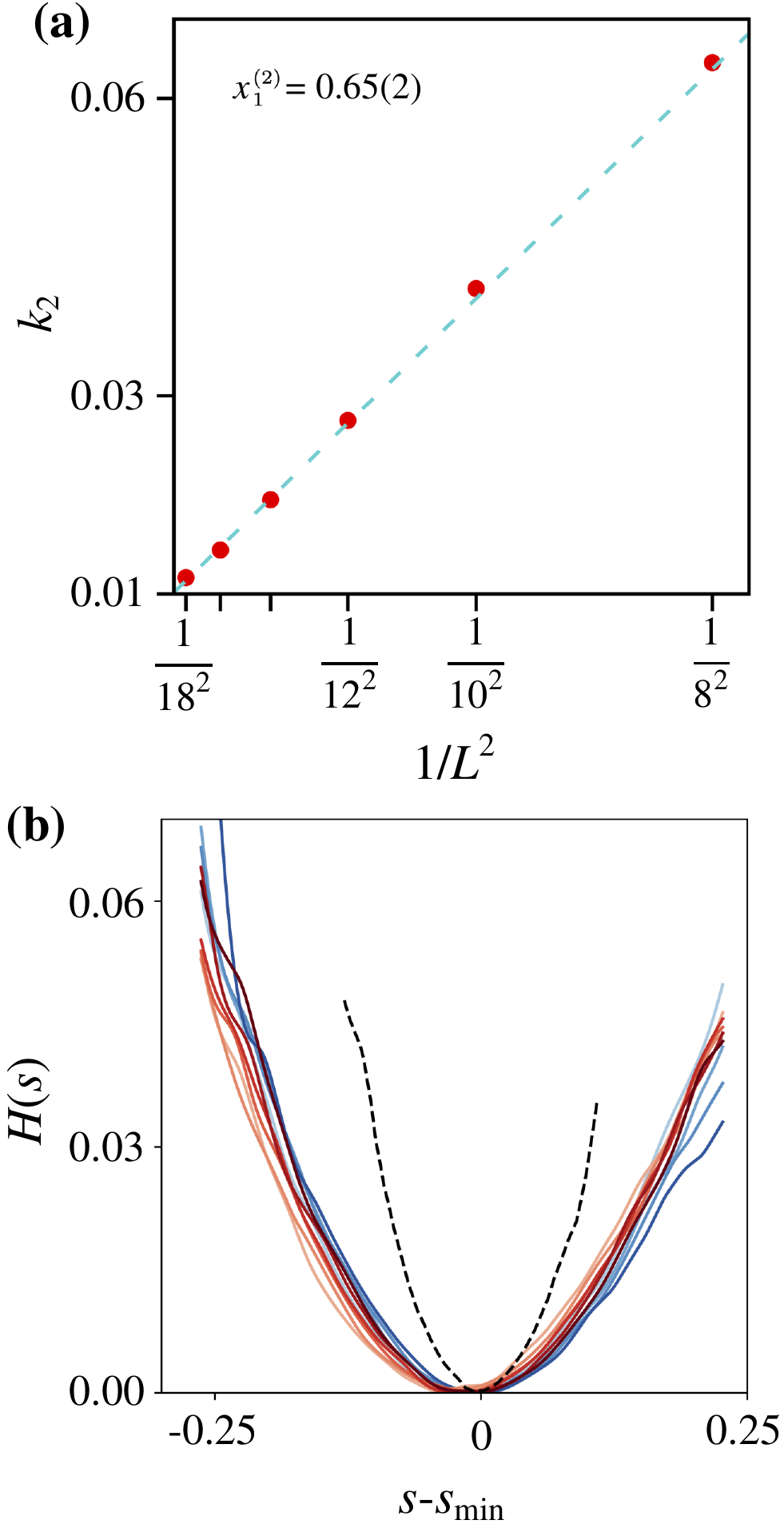}
  \caption{{\bf Multifractal scaling of the log-CFT}. Evidences of multifractality in (a) the 2nd cumulant $k_2$ vs $1/L^2$ giving a large $x_1^{(2)}=0.65(2)$ and (b) collapse of distribution of $Y(t)$ onto a single universal curve $H(s)$ for different values of $L$ and $t$ ($L=8 $ to $18$ at $t=16L$ by red lines and $t=11L$ to $27L$ for $L=18$ by blue lines). The universal curve $H(s)$ without U(1) symmetry, extracted from Ref.~\cite{Aidan_PRL}, is shown by the black dashed curve. The $x$-axis is plotted as $s-s_{\rm{min}}$ where the minimal value $s_{\rm{min}}=0.27(2)$ for U(1) symmetric case and $s_{\rm{min}}=0.14(2)$ without U(1) symmetry. The initial conditions are randomly chosen from all $\mathcal{Q}$ sectors. These results, together with the BKT nature of the sharpening transition, provide the strongest evidence that the universality class of the U(1) symmetric MIPT is unique and distinct from the Haar random case.
  } 
   \label{fig:multifractal}
   \end{figure}
  
In this subsection, we will probe the multifractal properties of the critical correlation function at the entanglement transition.

We know that at a generic critical point, where the system is scale-invariant, all moments of the correlation function vanish as a power law, 
\begin{equation}
    \mathcal{E}[C^n(r)] \sim \frac{B_n}{r^{2 x_1(n)}},
\end{equation}
where $\mathcal{E}[\dots]$ denotes averaging over different random samples.
If the scaling exponent $x_1(n)$ of the $n$th moment is a linear function of $n$, then the system is self-averaging, otherwise, it shows a multifractal scaling of the correlators. $x_1(n)$ can be obtained from the cumulant expansion,
\begin{eqnarray}
   \log \mathcal{E}[ C^n(r)] = && n \mathcal{E}[ \log C(r)] \nonumber \\
   &&+ \frac{n^2}{2!} \mathcal{E}[\{ \log C(r) - \mathcal{E}[ \log C(r)] \}^2 ]+\dots, \nonumber \\
\end{eqnarray}
which yields,
\begin{equation}
    x_1(n)= n x_1^{\rm{typ}} + \frac{n^2}{2!} x_1^{(2)} + \dots.
    \label{eq:x1n}
\end{equation}
Here, the coefficient of the linear term $x_1^{\rm{typ}}$ governs the decay of the average typical correlation function and finite higher cumulants $x_1^{(n)}$ represent the multifractal nature of the correlator. 
We computed $x_1^{\rm{typ}}$ from the difference between the average (over different trajectories) Lyapunov exponents $\lambda_1$ and $\lambda_0$ in the previous subsection. Here we compute the 2nd cumulant $k_2$ defined by, 
\begin{eqnarray}
   k_2=\frac{ \mathcal{E}[{\left( \lambda_1^{\vec{m}} -\lambda_0^{\vec{m}} \right)^2}] - \left( \mathcal{E}[{ \lambda_1^{\vec{m}} -\lambda_0^{\vec{m}}}] \right)^2 }{\alpha L} = 2\pi \frac{x_1^{(2)}}{L^2}.
\end{eqnarray}
Here, $\lambda_0^{\vec{m}}$ and $\lambda_1^{\vec{m}}$ are the Lyapunov exponents from each trajectory $\vec{m}$ in the circuit.
We numerically estimate $x_1^{(2)}= 0.65(2)$ from $k_2(L)$ vs $1/L^2$ shown in Fig.~\ref{fig:multifractal}(a). A large value of $x_1^{(2)}$ suggests the presence of strong multifractality.

This evidence is further verified by collapsing the distribution of $Y(t)= \alpha t ( \lambda_1^{\vec{m}} -\lambda_0^{\vec{m}}) $ onto the universal curve $H(s)$ 
which, satisfies  (up to a universal prefactor)~\cite{Aidan_PRL,multifractal1,multifractalLudwig},
\begin{equation}
    P[Y(t)]\sim \sqrt{\left( \frac{L}{2\pi \alpha t} \right)} \exp \left[- \frac{2\pi \alpha t}{L} H\left(\frac{2\pi \alpha t}{L} Y(t)\right) \right].
\end{equation}
Fig.~\ref{fig:multifractal}(b) shows the data collapse onto a single curve $H(s)$ vs $s$ for different values of $L$ and $t$ ($L=8 $ to $18$ at $t=16L$ by red lines and $t=11L$ to $27L$ for $L=18$ by blue lines). This  data collapse of $P[Y(t)]$ using a single curve $H(s)$ establishes multifractal scaling of correlation functions.
$H(s)$ shows a minimum at $s=s_{\rm{min}}$ given by,
\begin{equation}
    s_{\rm{min}}= \frac{d x_1(n)}{ dn}|_{n=0}=x_1^{\rm{typ}},
\end{equation}
In Fig.~\ref{fig:multifractal}(b), we shifted the x-axis to $s-s_{\rm{min}}$ where $H(s)$ shows a minimum near the typical scaling dimension $s_{\rm{min}}=x_1^{\rm{typ}} =0.27(2)$. The broadening of $H(s)$ is set by the variance $x_1^{(2)}=0.65(2)$.

We compare the multifractal spectrum $H(s)$ that we have obtained with the MIPT without any symmetry in Fig.~\ref{fig:multifractal} (b) (shown as a black dashed line). After shifting each curve by the location of the minimum in $H(s)$, $s_{\rm{min}}$, we are able to compare their shapes, which demonstrates that the higher average moments all differ as well. Thus providing strong evidence that the U(1) symmetry has modified the universallity class of the MIPT.

\section{Critical Properties of the Sharpening Transition}

\label{subsec:BKT}
\begin{figure*}[t!]
  \includegraphics[width=\textwidth]{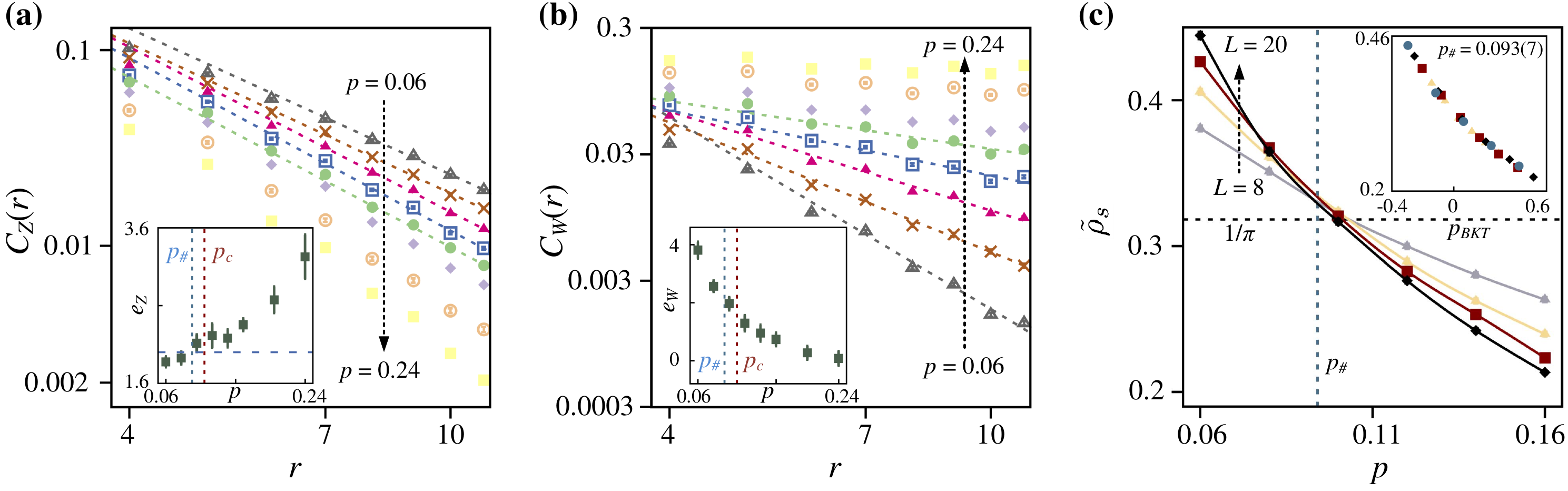}
  \caption{{\bf Charge sharpening from correlations}: (a) shows the spin-spin connected correlation function $C_z(r)$ vs $r$ for different $p$ values in a log-log plot. The decay is power-law for $p \lesssim 0.14$. For large $p $, the decay becomes faster than an algebraic decay. The inset shows the power-law slope, $e_z$ which remains $2$ for $p  \lesssim 0.14$. (b) shows the power-law decay of the string correlator $C_W(r)  \sim 1/|r|^{e_W} $ for smaller values of $p$ in a log-log plot. For large $p$, $C_W(r)$ becomes a constant with $e_W \rightarrow 0$ as shown in the inset. These features are consistent with a BKT criticality. (c) shows a finite-size estimate of $\rho_s$ extracted from $C_W$ with $p$ for different $L$. A finite-size BKT scaling collapse following Eq.\ref{eq:rho_scaling}, shown in the inset as a function of $p_{BKT}=(p-p_{\#})(\log L/a)^2$, yields $p_{\#}=0.093(7)$.  Last, we note that the stiffness $\tilde \rho_s$ appears to cross close to where the predicted universal value at the transition ought to be $\rho_s = \frac1{\pi}$ \cite{PRL_U1}.
}
   \label{fig:BKT}
  \end{figure*}
Motivated by the description of the charge sharpening transition in the infinite Hilbert space limit we analyze the sharpening transition in terms of observables that were shown to exhibit BKT  scaling \cite{PRL_U1}. In the large Hilbert space limit this is facilitated by mapping onto a weakly entangled model that can be simulated at large sizes with matrix product states. Here, however, we do not have such a mapping for qubit chains and are restricted to small system sizes where clearly identifying a BKT scaling unambiguously is close to impossible. Instead, here  our goal is to ascertain if the BKT sharpening diagnostics are both consistent with our numerical results on small sizes and look critical in a  window around the sharpening transition to see how it could affect the computed properties of the MIPT.

Therefore, we turn to numerically probe the charge-sharpening transition in a fixed charge density sector $\mathcal{Q}$ based on our finite-size numerics following Ref.\cite{PRL_U1}. We first compute the  average spin-spin connected correlation function,
\begin{equation}
    C_z(r)= \mathcal{E} \left [\langle \sigma^z_r \sigma^z_0 \rangle - \langle \sigma^z_r \rangle \langle \sigma^z_0 \rangle\right].
\end{equation}
Here, $\langle \dots \rangle$ denotes the expectation value of the operator in the quantum state of a trajectory and $\mathcal{E}[\dots]$ denotes averaging over different trajectories constituting the quantum circuit. The correlation function is expected \cite{PRL_U1} to go like $C_z(r)\sim \rho_s/r^2$ for $p\leq p_{\#}$
where $\rho_s$ is the superfluid stiffness and decay exponentially for $p>p_{\#}$.

We check for this behavior by  numerically calculating $C_z(r=L/2)$ at time $t=L$ for different system sizes $L=8$ to $22$ starting from initial conditions in $\mathcal{Q}=1/2$ sector. Fig.~\ref{fig:BKT}(a) shows $C_z(r)$ vs $r$ in a log-log plot for different values of $ 0.06 \leq p \leq 0.24$. For small values of $p \lesssim 0.14$, $C_z(r)$ shows a power-law decay while for larger values of $p \sim 0.24$, $C_z(r)$ decays faster than an algebraic decay and resembles  an exponential decay. A power-law fit $C_z(r) \approx 1/ r^{e_z}$ from the largest $4$ system sizes is shown by dashed lines and the slope $e_z$ vs $p$ is shown in the inset. The power-law function fits the data well for smaller $p$ values indicated by smaller error bars, while for larger values of $p$, the error-bar increases and the algebraic fit breaks down. Moreover, $e_z \approx 2$ over a range of values, $p \lesssim 0.14$ showing signatures of the BKT prediction of a critical charge fuzzy phase and a critical regime extending into the vicinity of our estimate of the MIPT.
Hence, this analysis strongly suggests that the  critical fluctuations of the charge sharpening transition still look critical at the measurement-induced transition $p_c$ on our available system sizes.

Next, we consider the string correlation function $C_W(r)$~\cite{PRL_U1} that is defined in terms of the string operator $W_{[0,r]}$, namely,
\begin{eqnarray}
    &&W_{[0,r]}=\prod_{0<i<r} \sigma^z_i, \nonumber \\
    &&C_W(r) = \mathcal{E} \left [\langle  W_{[0,r]} \rangle^2 \right].
\end{eqnarray}
$C_W(r)$ is expected to show a power-law decay in the charge fuzzy phase 
$C_W(r)  \sim 1/|r|^{2\pi \rho_s}$
and become independent of $r$ above the  sharpening transition.
We calculate the string correlator $C_W(r=L/2)$ at $t=L$ for $L=8$ to $22$ at different values of $p$ shown in Fig.~\ref{fig:BKT}(b). $C_W(r)$ shows a nice power-law decay for small $p$ values and becomes a constant for larger $p$ values. Hence, this numerical result is qualitatively in good agreement with the expectation from the BKT prediction. A power law fit $C_W(r) \sim 1/|r|^{e_W}$ (shown by the dashed lines) yields the slope $e_W$ which is shown as a function of $p$ in the inset. The fitted power-law $e_W$ monotonically decreases with increasing $p$ as expected in any finite-size analysis. This prediction also suggests that these string correlations remain critical up to our estimate of the entanglement transition as they appear within a given sector.

In order to provide an additional  test of the BKT transition, we introduce a finite size scaling function for the  stiffness $\tilde\rho_s(p,L)$. From the expectation $C_W(r)  \sim 1/|r|^{2\pi \rho_s}$ we construct,
\begin{eqnarray}
    \tilde\rho_s(p,L) = - \frac{\log C_W(L/2)}{2 \pi \log (L/2)},
\end{eqnarray}
which shows a crossing for different system sizes $L=8$ to $20$ as a function of $p$ as depicted in Fig.~\ref{fig:BKT}(c). From the   crossing we estimate the location of the sharpening transition $p_{\#}$ and perform a finite-size BKT scaling analysis from the ansatz,
\begin{equation}
    \tilde{\rho}_s \sim h_s[(p-p_{\#}) (\log L/a)^2],
    \label{eq:rho_scaling}
\end{equation}
where $h_s$ is an unkown scaling  function with $p_{\#}$ and $a$ as the fitting parameter shown in the inset. Our analysis gives $p_{\#} =0.093(7)$, which is very close to our previous estimates of the sharpening transition (e.g. in Fig.~\ref{fig:method}). Moreover, the value of the stiffness is expected to undergo a quantized jump if the sharpening transition is of the predicted BKT type of size $1/\pi$ \cite{PRL_U1}. Based on our estimate of the crossing in $\tilde \rho_s$ at $p_{\#}$ we find good agreement with the stiffness jumping by $1/\pi$  within the error bars of our estimate of the sharpening transition.
All of this together provides strong evidence that the sharpening transition in qubit chains is of the BKT type.

In the following subsection we discuss a possible interpretation of the entanglement critical properties that we have found with values that are anomalously large relative to measurement-induced criticality without any symmetry. Following the perspective that our finite size estimates of the critical properties are going to ``feel'' the sharpening transition on finite size simulations, due to its close proximity as depicted in the finite size cross-over diagram in Fig.~\ref{fig:BKT_illustration}, we are able to provide a clear explanation of our results as being shifted by the critical properties of a BKT sharpening transition.

\section{Interpretation of the estimated critical entanglement properties}\label{sec:illustration}

Our numerical results presented in Sec.\ref{sec:phases} suggest that the two critical points, the charge-sharpening and the entanglement transition, are approaching to each other as we go to $\mathcal{Q}$ sectors away from half-filling. Hence, the entanglement critical properties, which are of our main interests in the present work, may be influenced by the nearby charge-sharpening transition. To ascertain this more carefully, in the previous section we showed that on our finite size simulation within a single sector, the charge sharpening correlations remain critical at the entanglement transition. Here we pose two plausible interpretations of our above (anomalously large) estimates of the critical exponents of MIPT considering the nearby sharpening criticality. Despite the scenario, as the sharpening transition is BKT, we reiterate that we find strong evidence (through the multifractal scaling) that the U(1) symmetric MIPT belongs to a unique universality class.

\begin{figure}[t!]
\centering
  \includegraphics[width=0.75\columnwidth]{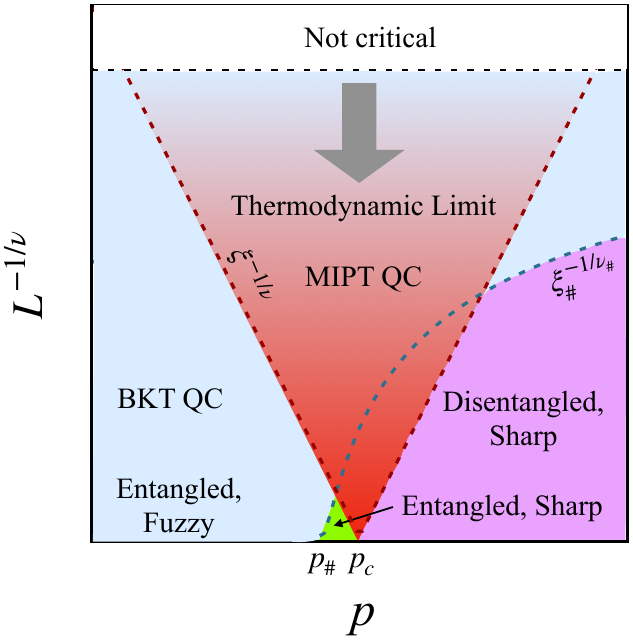}
  \caption{{\bf Schematic finite-size cross over diagram showing the two nearby transitions}. 
  The system undergoes a charge sharpening transition from the charge fuzzy to sharp phase at the measurement rate $p=p_{\#}$. This is followed by the usual measurement induced phase transition from volume-law entanglement to area-law phase at $p=p_c$. These three phases, namely the entangled-fuzzy, entangled-sharp and the disentangled-sharp, are marked in the figure. The red and the blue dashed lines show the correlation lengths $\xi$ and $\xi_{\#}$ corresponding to the MIPT and the BKT-sharpening transition respectively. The blue region shows the finite-size BKT critical fan which spans over the entire charge-fuzzy phase and bounded by $\xi_{\#}$ from the charge-sharp phase. The red region shows the finite-size MIPT critical fan bounded by $\xi$.
  The sharpening criticality has a broad critical fan, intertwining these two transitions at smaller system sizes accessible in finite-size numerics. 
  }
   \label{fig:BKT_illustration}
  \end{figure}

(a) The first scenario is that the MIPT and charge-sharpening transition occur at distinct critical measurement rates (i.e. separate locations)  in the thermodynamic limit. But at finite accessible system sizes, the close proximity of the BKT-sharpening criticality and the entanglement transition leads us to interpret the entanglement critical exponents listed in table \ref{tab:exponents} as receiving a contribution from the critical sharpening fluctuations. A scenario for how this fits into the finite size scaling regime we access is depicted in the schematic diagram for the two transitions in Fig.~\ref{fig:BKT_illustration}. The finite-size critical fans of MIPT and charge-sharpening transition shown by red and blue colored regions respectively bounded by their corresponding correlation lengths $\xi$ and $\xi_{\#}$, (shown by dashed lines) overlap. In particular, the finite-size estimation of the critical exponents in the 2nd column of table \ref{tab:exponents} can be reinterpreted as receiving a combined contribution from the two transitions: their known values in the BKT universality class (denoted by the subscript $\#$ in 4th column) and the MIPT with a U(1) symmetry (denoted by the subscript $E$ obtained after subtracting the corresponding BKT values from the 2nd column) and we compare these values (in magnitude) to the MIPT transition without any symmetry (denoted by the subscript $H$ in the 3rd column).

First, we discuss if the presence of the nearby sharpening transition affects our estimate of the exponent $\eta$. 
Due to the close proximity of the charge-sharpening transition with the entanglement transition and the fact that the sharpening fluctuations can be critical within a given sector, the scaling dimension $\eta(L)$ we pick up at finite sizes should come from the combination of the measurement-induced transition and the sharpening transition, i.e. $\eta(L)=\eta_{\#}+\eta_{E}$. Here $\eta_{\#}=1/4$ is assumed for the BKT charge sharpening transition,  $\eta_E$ is the critical exponent of the U(1) symmetric entanglement transition (once we subtracted the BKT values from the finite-size estimation $\eta$), and we denote the finite size estimate as $\eta(L)$ as we do not expect that this estimate necessarily holds in the thermodynamic limit. As such, this would provide the estimate of $\eta_E = 0.32(3)$, which is close to  $\eta_H=0.26(1)$ from the Haar random problem without symmetry.

Second, we turn to our estimate of the effective central charge $c_{\mathrm{eff}}$.
The BKT sharpening transition is also a CFT (but it is not multifractal) and has a well-defined central charge $c_{\#}=1$. Our analysis of the free energy will then acquire a shift from these critical modes and we therefore expect that our estimate of the effective central charge is offset $c_{\mathrm{eff}}(L)=c_{\#}+c_{\mathrm{eff},E}$, which yields a $c_{E, \mathrm{eff}}=0.27(1)$ that is quite close to  the transition without symmetry $c_{\mathrm{eff},H}=0.25(2)$, and again we denote $c_{\mathrm{eff}}(L)$ as a finite size estimate that does not necessarily hold in the thermodynamic limit. Thus, the interpretation of our estimate of the effective central charge being contaminated the nearby sharpening transition provides a natural explanation of our large estimate. Conversely, it also provides additional evidence that the sharpening transition is indeed a BKT transition. 

To conclude, we summarize the implication of Fig.~\ref{fig:BKT_illustration} to our estimates in the thermodynamic limit.  We expect that as we approach $L\rightarrow \infty$, the sharpening transition separates from MIPT in $p$ space and at $p=p_c$ only the entanglement transition contributes to the exponents that implies
$\eta(L)\rightarrow \eta_E$ and $c_{\mathrm{eff}}(L)\rightarrow c_{\mathrm{eff},E}$.

Last, we turn to the multifractal properties of the transition.
As the BKT transition is not multifractal, it 
ensures that its contribution to $x_1(n)$ in Eq.~\eqref{eq:x1n} is linear in $n$. 
Combining this with Eq.~\eqref{eq:x1n}, we see that the only possible effect of the nearby BKT transition on the multifractal scaling dimension is to shift the location of the minimum of $H(s)$ to $s_{\rm{min}}= \frac{ \eta(L)}{2}.$
Using our numerical estimate of $s_{\rm{min}} \approx 0.27(2)$ from Fig.~\ref{fig:CFT}, we get an estimate $\eta_E= 0.3(3)$ This is consistent with our understanding from section~\ref{subsec:BKT}. Moreover, all the higher cumulants i.e., $x_1^{(n)}$ are unaffected by the BKT criticality and hence solely characterize the entanglement universality class. Our observation of $x_1^{(2)}=0.65(2)$ being significantly different from that of the MIPT without symmetry, $x_1^{(2)}=0.15(2)$, thus strongly suggests that the U(1) MIPT is characterized by a unique multifractal log-CFT that is distinct from its non-symmetric counterpart.

(b) An alternative possible scenario is that the MIPT and charge-sharpening transitions occur at the same measurement rate (i.e. location in parameter space and are inseparable even in the thermodynamic limit). In this case, the critical exponents of this transition are anomalously large compared to non-symmetric (usual Haar) MIPT. This again suggests presence of $U(1)$ symmetry changes the universality class of MIPT. However, this scenario seems less likely as it is in direct contradiction with  analytic results obtained in a suitably defined infinte on-site Hilbert stape limit ($d\rightarrow \infty$), which finds that the entanglement transition and charge-sharpening transition are separate~\cite{PRX_U1}. 

\section{Conclusion}

In conclusion, this study has provided a comprehensive analysis of the universality class of the measurement-induced phase transition (MIPT) in U(1) conserving hybrid quantum circuits. Our sector-resolved approach enabled us to investigate charges far from the middle sector and access system sizes up to $L = 48$ for a conserved charge density of $\mathcal{Q} = Q/L = 1/12$. Our findings demonstrate that each sector exhibits Lorentz invariance with a phase boundary that follows the variance of the binomial distribution $p_c(\mathcal Q) \propto \mathcal Q (1-\mathcal Q)$.
The MIPT is characterized by a logarithmic conformal field theory, which we quantified through the examination of the transfer matrix and its Lyapunov spectrum governing the quantum evolution.
The effective central charge ($c_{\mathrm{eff}}$), the anomalous scaling dimension ($\eta$), and the multifractal scaling of correlation functions in the conformal field theory were all found to be distinctly different from the MIPT without a conservation law. We then analyzed the charge sharpening physics and found it to be consistent with a BKT transition though our small system size study is not conclusive. Importantly, we also found that even within a single charge sector, critical sharpening physics can still appear in the dynamics of correlation functions averaged over measurement outcomes and these still look critical up to the MIPT. Thus, we provided a natural interpretation to our results that they have been shifted by BKT-related critical exponents. A separate scenario, which seems less likely, is that the sharpening transition and the entanglement transition coincide. However, this is at odds with controlled calculations in the limit of an infinite onsite Hilbert space dimension and we expect this can be resolved in future studies that find models that ``pull apart'' the locations of the sharpening and entanglement transitions. Nonetheless, regardless of this, by studying the multifractal spectrum of the log-CFT at the entanglement transition our work  offers strong numerical evidence that the U(1) conservation law alters the universal nature of the MIPT.

\acknowledgements{We thank Srivatsan Chakram, David Huse, and Matthew Fisher for insightful discussions as well as Utkarsh Agrawal, Sarang Gopalakrishnan, Andrew Potter, and Romain Vasseur for discussions and collaborations on related work. This work was partially supported by the Abrahams Postdoctoral Fellowship at the Center for Materials Theory Rutgers (A.C.), the Army Research Office Grant No.~W911NF-23-1-0144 (J.H.P.) and a Sloan Research Fellowship (J.H.P.). 
J.H.W.\ acknowledges financial support from NSF grant DMR-2238895.
This work was performed in part  at the Aspen Center for Physics, which is supported by National Science Foundation grant PHY-2210452 (J.H.W., J.H.P.)
The computations were performed using the Beowulf cluster at the Department of Physics and Astronomy of Rutgers University; and the Office of Advanced Research Computing (OARC) at Rutgers, The State University of New Jersey (http://oarc.rutgers.edu), for providing access to the Amarel cluster. The Flatiron Institute is a division of the Simons Foundation.}

\appendix
\begin{widetext}
    
\begin{figure}[h!]
  \includegraphics[width=0.99\textwidth]{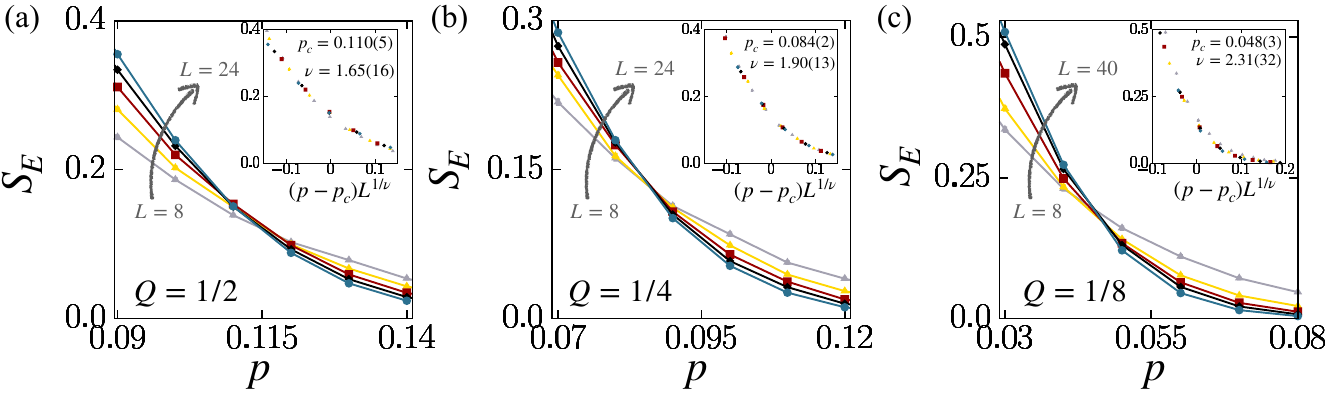}
  \caption{Entanglement entropy of the ancilla $S_E$  vs $p$ at (a) $Q=L/2$ for $L=8$ to $24$, (b) $Q=L/4$ for $L=8$ to $24$ and (c) $Q=L/8$ for $L=8$ to $40$. $S_E$ is collapsed with the scaling form Eq.~\eqref{eq:scaleall} (and also consistent with Eq.~\eqref{eq:scaleKun}) to obtain $p_c(\mathcal{Q})$ and $\nu(\mathcal{Q})$ shown in the insets.}
   \label{fig:SM_anc}
  \end{figure}  
  
\section{Critical exponents from power-law scaling of the entangle order parameter} \label{secSM:OP}
To probe the entanglement critical point, we use a ``single-site Ancilla" method to calculate the order parameter $S_{E}$. After coupling the ancilla at $t=0$ in a fixed charge sector $\mathcal{Q}=Q/L$, we evolve the U(1) symmetric circuit with measurements upto $t=2 L$ to reach the steady state. We then compute the entanglement entropy $S_{E}$ of the ancilla for different measurement rates $p$ and different system sizes $L$ commensurate with Q. Fig.~\ref{fig:SM_anc}(a)-(c) shows the variation of $S_{E}$ with $p$ in three different charge sectors (for $n=1$), $\mathcal{Q}=1/2,1/4$ and $1/8$. We have shown $S_{E}$ vs $p$ in the main text for $\mathcal{Q}=1/12$. In each cases, with increasing $p$, $S_{E}$ decreases from a volume law phase to an area law phase crossing at $p=p_c$ between different $L$s. The critical point is obtained using a finite size scaling analysis (at a fixed time) with two different protocols: (i) using the standard finite-size scaling ansatz for a second order transition,
\begin{equation}
    S_{E}(L;\mathcal Q)\sim \tilde{h}_{\mathcal Q}[(p-p_c(\mathcal Q))L^{1/\nu(\mathcal Q)}]
    \label{eq:scaleall}
\end{equation}
and collapsing the data onto a universal curve $\tilde{h}_{\mathcal Q}$ with two parameters $p_c(\mathcal Q)$ and $\nu(\mathcal Q)$ in each $\mathcal{Q}$ sector. (ii) We use the ansatz proposed in Ref.\onlinecite{PRX_U1},
\begin{equation}
    S_{E}(L;\mathcal Q)\sim \bar{h}_{\mathcal Q}[(p-p_c(\mathcal Q))L^{1/\nu(\mathcal Q)}, y(\mathcal Q) L^{-\omega(\mathcal Q)}],
\end{equation}
 
where the leading irrelevant scaling variable $y$ is incorporated to take into account the shift of the crossing points with increasing system size. $\bar{h}_{\mathcal Q}$ can be expanded in Taylor series around $p=p_c$ as,
\begin{equation}
    S_{E}(L;\mathcal Q) = a+ b (p-p_c(\mathcal Q))L^{1/\nu(\mathcal Q)} + c (p-p_c(\mathcal Q))^2L^{2/\nu(\mathcal Q)} + d /L^{\omega(\mathcal Q)}.
    \label{eq:scaleKun}
\end{equation}
We numerically find the exponents $p_c(\mathcal Q)$ and $\nu(\mathcal Q)$ and the fitting parameters $a,b,c,d$ (omitted the $\mathcal Q$ dependence for simplicity of the notation) by non-linear fitting from the data. The values $p_c(\mathcal Q)$ and $\nu(\mathcal Q)$ shown in the main text in Figs.~\ref{fig:method}(c) and \ref{fig:ancilla} (b) as well as the insets of Fig.~\ref{fig:SM_anc} are consistent with both the scaling anstaz within errorbars.

We next show how we extract the dynamical critical exponent $Z$. We collapse $ S_{E}$ with $t/L^{z(\mathcal{Q})}$ for different $L$s in a fixed $\mathcal{Q}$ sector. Fig.~\ref{fig:SM_zexp}(a)-(c) show the data collapse for $\mathcal{Q}=1/2,1/4$ and $1/8$ ($\mathcal{Q}=1/12$ is shown in the main text). This shows $z\approx 1$ across $\mathcal{Q}$ sectors as shown in Fig2(c) of the main text. This justifies our choice of the aspect ratio $t/L = 2$ (i.e. an $\mathcal{O}(1)$ number) in finding $p_c$ and $\nu$. 
  
\begin{figure}[t!]
  \includegraphics[width=0.85\textwidth]{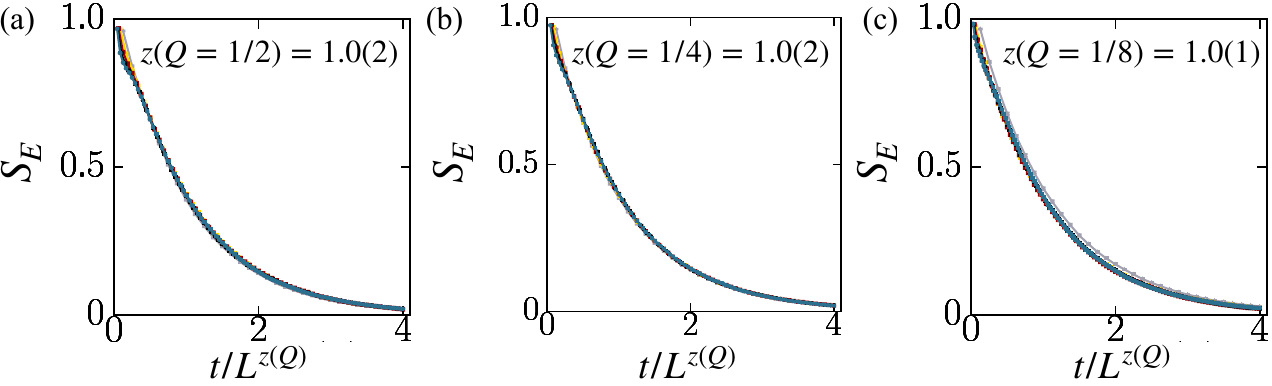}
  \caption{Entanglement entropy of the ancilla $S_E$  is collapsed with re-scaled time $t/L^{z(\mathcal{Q})}$ following Eq.\ref{eq:SE} at (a) $Q=L/2$ for $L=8$ to $24$, (b) $Q=L/4$ for $L=8$ to $24$ and (c) $Q=L/8$ for $L=8$ to $40$. This gives the dynamical exponent $z\approx 1$ for all charge sectors.  }
   \label{fig:SM_zexp}
  \end{figure}

\section{BKT finite-size scaling analysis of the charge-sharpening phase transition} \label{secSM:BKT}

\begin{figure}[h!]
  \includegraphics[width=0.99\textwidth]{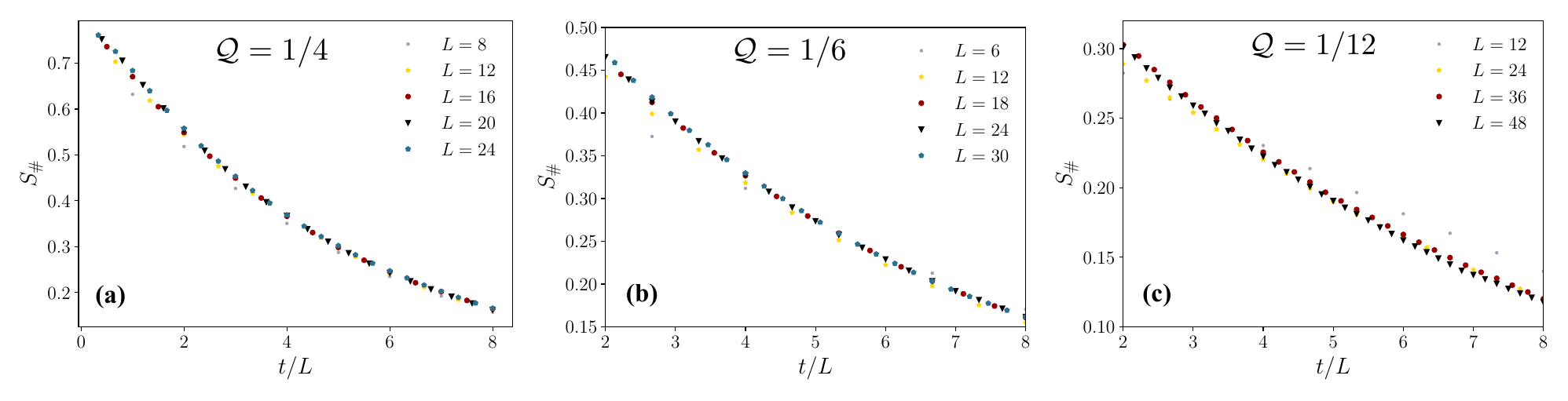}
  \caption{Charge-sharpening order parameter $S_\#$ of various system sizes $L$ is collapsed with recaled time $t/L$ at (a) $Q=L/4$ (b) $Q=L/6$ and (c) $Q=L/12$. The collapse at the charge-sharpening phase transition point indicates the dynamical critical exponent $z\approx 1$ for all charge sectors.  }
   \label{fig:charge_sharpening_z}
  \end{figure}  

To understand the nature of the charge-sharpening phase transition in a quantum circuit system, we employed the ancilla method as outlined in Ref.\cite{PRX_U1}. This method involves coupling an ancilla to two distinct charge sectors, represented as $|\Psi\rangle=\left|\psi_{{Q}}\right\rangle|\Uparrow\rangle+$ $\left|\psi_{{Q}+1}\right\rangle|\Downarrow\rangle$; where $\left|\psi_{Q}\right\rangle$ symbolizes states within the charge sector ${Q}$, while $|\Uparrow \rangle$ and $|\Downarrow\rangle$ represent ancilla states. The order parameter $S_{\#}$, signifying the von Neumann entanglement entropy of the reduced density matrix ${\operatorname{Tr}}_{\text{ancilla}} |\Psi\rangle \langle\Psi|$, is used to probe the charge-sharpening phase transition. To compute the entropy $S_{\#}$, we adjusted the measurement rates $p$ and system sizes $L$ to be commensurate with Q. Fig.~\ref{fig:charge_sharpening_collapse} illustrates the variation of $S_\#$ with $p$ for different charge sectors, $\mathcal{Q}=1/2, 1/4, 1/6$ and $1/12$ at fixed time $t=2L$.

We used finite-size scaling laws of quantum criticality to understand the behavior near the charge-sharpening phase transition. For our analysis, we tested two possible types of transitions: a second-order phase transition and a BKT transition. For the former, the scaling ansatz matched that for the entanglement phase transition as seen in Eq.~\eqref{eq:scaleKun}. On the other hand, for the BKT transition, we employed a different scaling ansatz:
\begin{equation}
S_{\#}(L;\mathcal Q)\sim g_{\mathcal Q}[(p-p_{\#}(\mathcal Q))(\log L/a)^2, y_{\#}(\mathcal Q) L^{-\omega_{\#}(\mathcal Q)}],
\end{equation}
Here, $a$ is the ultraviolet scale of BKT criticality, $y_{\#}$ signifies the amplitude of the subleading term with the scaling dimension $\omega_{\#}(\mathcal Q)$, and $L$ represents the system size. The choice of logarithmic scaling for $L$ is guided by the BKT scaling law for the correlation length $\xi/a \sim \exp(1/\sqrt{|p-p_c|})$. In the vicinity of the critical point, $g_{\mathcal Q}$ can be perturbatively expanded,
\begin{equation}
\label{eq:BKT_scaling}
S_{\#}(L;\mathcal Q)\sim a_{\#}+b_{\#}(p-p_{\#}(\mathcal Q))(\log L/a)^2+c_{\#}(p-p_{\#}(\mathcal Q))^2(\log L/a)^4 + d_{\#} L^{-\omega(\mathcal Q)},
\end{equation}
where the fitting parameters are $p_{\#}$, $a$ and $a_{\#}, b_{\#}, c_{\#}, d_{\#}$. We tested a wide range of the exponent $\omega_{\#}$ to establish the error bars. 

We next discuss the dynamics of the charge-sharpening order parameter $S_\#(t)$ at fixed measurement rate $p=p_{\#}$. As shown in Fig. \ref{fig:charge_sharpening_z}, we collapse $S_\#$ with the rescaled time $t/L$ for different system sizes. We observe good collapse for different sectors $\mathcal{Q}=1/4$, $1/8$ and $1/12$, indicating the dynamical critical exponent $z=1$ for all charge sectors at the charge-sharpening phase transition point.

\section{Two methods to calculate mutual information between a pair of ancilla} \label{secSM:eta}
In the calculation of the mutual information between a pair of ancillas, here we compare the newly developed ``single-site ancilla" method explained in Sec.\ref{subsec:singleanc} of the main text with the previously used ``bond-ancilla" method (to find $p_C$ and $\nu$) in Ref.\onlinecite{PRX_U1}. 
In the ``bond-ancilla" method, to probe the entanglement phase transition in a fixed global charge sector $\mathcal{Q}$, an ancilla is entangled in a Bell state with a bond of the adjacent sites $j$ and $j+1$ in the system yielding the wave function $ |\tilde{\Psi} \rangle =(1/\sqrt{2}) \big[|\dots \uparrow\downarrow \dots \rangle |\Uparrow \rangle +|\dots \downarrow\uparrow \dots \rangle |\Downarrow \rangle \big]$. Here the states $|\dots \uparrow\downarrow \dots \rangle $ and $|\dots \downarrow\uparrow \dots \rangle$ belong to same global charge sector $\mathcal{Q}$ and are orthogonal due to the orthogonal configuration at the bond. Although this method of entangling ancilla accurately predicts the entanglement critical point ($p_c$ and $\nu$), it is difficult to extend this method to calculate mutual information between a pair of ancillas and associated $\eta$ exponents at $p_c$. As we know, correlation function \cite{AidanPRB} is obtained from the mutual information of the two ancilla entangled to the system. Entangling two ancillas each coupled to a bond requires $4$ projection operations on the wave function (ex: $|\dots \uparrow\downarrow \dots \downarrow \uparrow \dots \rangle |\Uparrow\Downarrow\rangle$ ) and makes this method numerically challenging due to collapse of the wavefunction. Moreover, if we go to the charge sectors away from the middle sectors, the dimension of the constrained Hilbert-space (due to U(1) symmetry) reduces and it becomes increasingly difficult to couple two ancillas to two orthogonal states without collapsing the system wavefunction.

To circumvent these issues, we used the efficient ``single-site ancilla" method to couple the ancilla and calculated correlation functions at different charge sectors. We compare the two methods (the ``single-site ancilla" and the ``bond-ancilla") in Fig.~\ref{fig:SM_bondsite} (a)-(b)at the middle sector where the $\eta$ exponent (for $n=1$) extracted from both the methods are in good agreement. This establishes the consistency of the new method. However, the ``bond-ancilla" method requires running $\sim 1.5$ times more samples compared to the ``single-site ancilla" method to obtain similar size of the statistical ensemble over which we averaged the correlation function. Moreover, the ``bond-ancilla" method required us to wait longer to reach the steady state before coupling the ancillas ($t_0 \sim L^2$), while for the ``single-site ancilla" method it is enough to wait till $t_0 \sim 2 L$. In Fig.~\ref{fig:SM_bondsite}, we showed both (a) and (b) at $t_0=2L^2$ for the purpose of comparison.

Next, we use the ``single-site ancilla" method to calculate the correlation function at $\mathcal{Q}=1/8$. Fig.~\ref{fig:SM_bondsite}(c) shows the scaling collapse with $\eta=0.69(5)$. We note that, even after using the efficient method at $\mathcal{Q}=1/8$, a large number of samples collapsed due to the above-mentioned post-selection problem and the size of the statistical ensemble we averaged, is only $\sim 1/4$ compared to that at the middle-sector. This leads to poor quality of the scaling collapse with large error bars at $\mathcal{Q}=1/8$ compared to that at half-filling (Fig.~\ref{fig:SM_bondsite}(b)).

\begin{figure}[t!]
\includegraphics[width=0.85\textwidth]{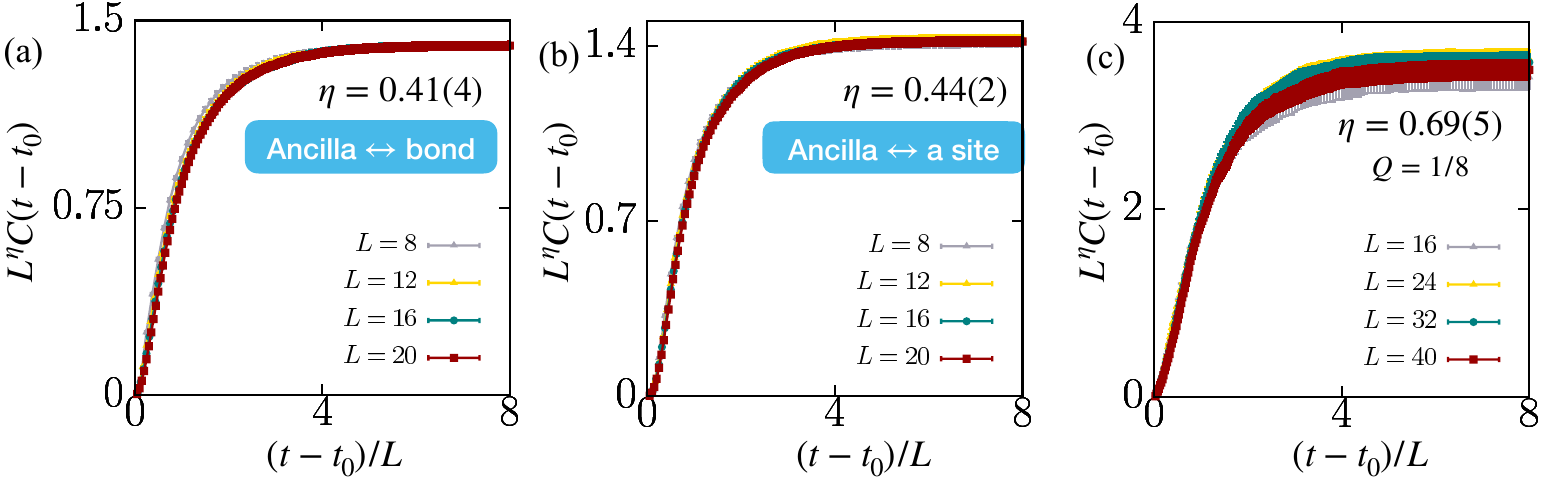}
  \caption{The scaling collapse of the correlation function with time using Eq.\ref{eq:eta}: the two ancillas are coupled using the ``bond-ancilla" method in (a) and using the ``single-site ancilla" method in (b). The critical exponent $\eta$ extracted using these two methods are $\eta=0.41(4)$ from the ``bond-ancilla" method and $\eta=0.44(2)$ from the ``single-site ancilla" method at $p_c=0.110$ and $\mathcal{Q}=1/2$. These values are in good agreement with each other, establishing consistency between these two methods. (c) shows the scaling collapse of the mutual information at $\mathcal{Q}=1/8$ (using the ``single-site ancilla" method) and $p_c= 0.048$. This gives $\eta=0.69(5)$, but with a poorer quality of collapse due to averaging over smaller number of samples. We use $n=1$ for this plot.}
   \label{fig:SM_bondsite}
  \end{figure}

  \section{Computation of the anisotropy parameter $\alpha$}
  \label{sec:alpha}
  To compute the anisotropy parameter $\alpha$ in the area of the cylinder of the CFT $A=\alpha L t$, we followed the procedure prescribed in Ref.\cite{Aidan_PRL}. We compute two kinds of correlation functions: (i) the ``space-like" correlation function $I_{\rm{space}}=I_{n=1}(A,B:\delta r=L/2,\delta t =0)$ from the mutual information of the two ancillas $A$ and $B$, coupled in the steady state of the circuit at the same instant of time $t_1=t_2=t_0(=2L)$, at a spatial separation $|r_2-r_1|=L/2$. (ii) the ``time-like" correlation function $I_{\rm{time}}(\delta t)=I_{n=1}(A,B:\delta r=0,\delta t)$ from the mutual information of the two ancillas, where the first ancilla is coupled at $t_1=t_0=(2L)$ and the second one is coupled at a separation in time $t_2-t_1=\delta t$, at the same site in the system. After coulping the two ancillas in the circuit, we calculate the time evolution of the correlation functions $I_{n=1}(A,B)$ with time $t-t_2$. This is plotted in Fig.~\ref{fig:SM_alpha}(a) where the initial conditions are chosen only from the middle sector $\mathcal{Q}=1/2$ for a $L=16$ site system. We set the measurement rate at $p=p_c=0.110$ at half-filling. Here, $I_{\rm{space}}$ is shown by the yellow line while $I_{\rm{time}}$ is shown for two values of $\delta t =8$ and $9$ which cross $I_{\rm{space}}$. 
  \begin{figure}[h!]
  \includegraphics[width=0.60\textwidth]{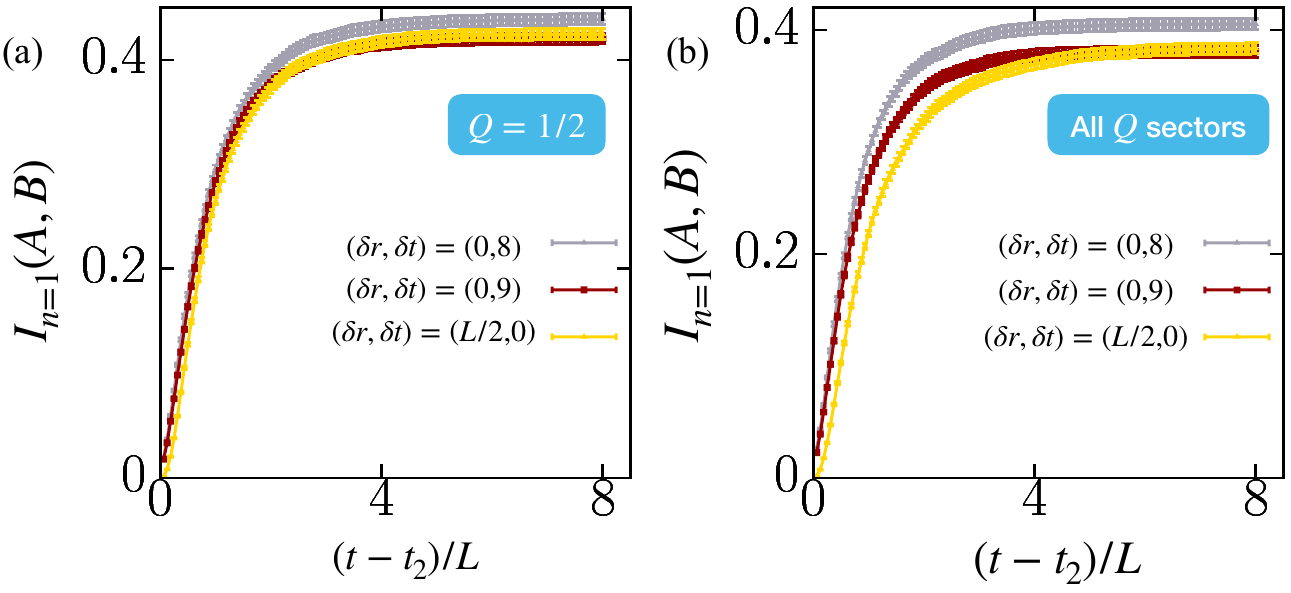}
  \caption{Computation of the anisotropy paremeter $\alpha$: at $p=p_c(\mathcal{Q}=1/2)=0.11$, we plot the ``space-like" correlator $I_{\rm{space}}$ with yellow color and the ``time-like" correlators $I_{\rm{time}}(\delta t=8)$ (grey) and $I_{\rm{time}}(\delta t=9)$ (red) for two values of $\delta t$ which cross $I_{\rm{space}}$ at late times. We obtain $\alpha =0.51(3)$ in case (a), when we restrict to only $\mathcal{Q}=1/2$ trajectories and this matches with the case (b) with $\alpha=0.50(1)$ where the initial conditions are randomly chosen over all charge sectors.   }
   \label{fig:SM_alpha}
  \end{figure}
From this, we have to find out the value of $\delta t=t_{*}$ at which the two types of the correlations match, $I_{\rm{space}}=I_{\rm{time}}(t_*)$. Since, numerically we can only calculate $I_{\rm{time}}(\delta t)$ on a grid of $\delta t$ with grid size$=1$, we interpolate to obtain,
  \begin{equation}
      t_* = 8 + \frac{I_{\rm{space}}-I_{\rm{time}}(8)}{I_{\rm{time}}(9)-I_{\rm{time}}(8)}(9-8). 
  \end{equation}
This gives $t_*=8.768$. We compute the value of $\alpha$ from $\alpha=\log (1+\sqrt{2})L/(\pi t_*)$ derived in Ref.\cite{Aidan_PRL}. This gives $\alpha= 0.51 (3)$ at half-filling. We use this value of $\alpha$ for numerical results in Sec.\ref{sec:cft}.

Furthermore, we also compute $\alpha$ when the system is initialized to states superposing with all charge sectors. This is shown in Fig.~\ref{fig:SM_alpha}(b) at $p=p_c=0.110$ for $L=16$. Following similar procedure explained above, we obtain $t_*=8.912$ which yields $\alpha=0.50(1)$. Hence the estimates of $\alpha$ restricting to middle sectors and that from all charge sectors are in good agreement. This is also consistent with the fact that for large $L$ (in thermodynamic limit), the middle sectors ($\mathcal{Q}=1/2$) dominates over all other charge sectors as shown in Fig.~\ref{fig:CFT}(b).

\section{Dependence of initial conditions on $f_1(L)$} \label{secSM:f1}
In section \ref{sec:cft} of the main text, we showed the dependence on the initial conditions of the free energy $f(L)$ and the difference between the generalized free energies $f_1(L)-f(L)$. We discussed that when we restrict to the contribution only from a fixed charge sector, the free energy $f^{\mathcal{Q}}(L)$ picks up a leading $1/L$ dependence (in addition to the usual $1/L^2$ scaling). Here we show that the generalized free energy $f_1(L)$ shows a similar $1/L$ scaling when we compute $f_1^{\mathcal{Q}}(L)$ from the trajectories projected to $\mathcal{Q}$ sector at late times. This is shown in Fig.~\ref{fig:SM_f1} by blue triangles at $\mathcal{Q}=1/2$ and $p=p_c=0.11$. $f_1^{\mathcal{Q}=1/2}(L)$ can be fit well using a form,
  \begin{figure}[t!]
  \includegraphics[width=0.28\textwidth]{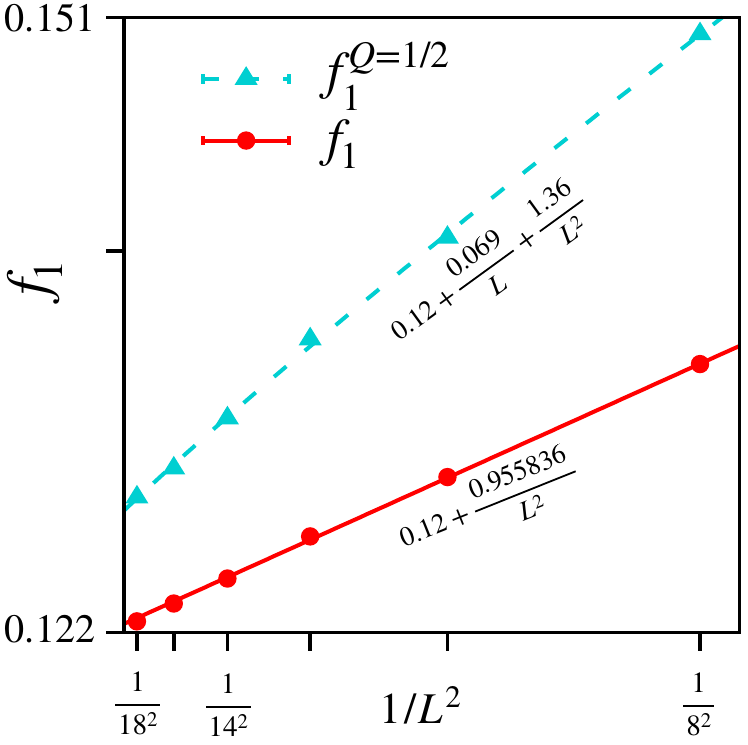}
  \caption{The generalized free energy $f_1(L)$ is shown as a function of $1/L^2$ for two different initial conditions: the blue triangles $f_1^{\mathcal{Q}=1/2}$ correspond to restricting to the trajectories to $\mathcal{Q}=1/2$ charge sector while the red circles correspond to the contributions from all trajectories with unrestricted charge. $f_1^{\mathcal{Q}=1/2}$ can be fit well by adding a $1/L$ dependence to the usual $1/L^2$ scaling form that applies to the unrestricted charge case. }
   \label{fig:SM_f1}
  \end{figure}

 \begin{equation}
f_1^{\mathcal Q}=f_1^{\mathcal Q}(L\rightarrow \infty)+ a' \mathcal Q/ L+ \frac{a''}{L^2}.
\end{equation}
 This $1/L$ dependence coming from the chemical potential fixing the density to that charge sector is cancelled in the difference $f_1^{\mathcal{Q}}(L)- f^{\mathcal{Q}}(L)$ from each trajectory with charge $Q$. This is verified in Fig.~\ref{fig:CFT}(b) of the main text. We also plot $f_1(L)$ from all trajectories with unrestricted charge in Fig.~\ref{fig:SM_f1} by the red circles. In this case, absence of chemical potential leads to the usual $1/L^2$ scaling. 
  \end{widetext}

\bibliography{main.bib}
\end{document}